\documentclass[12pt]{article}

\usepackage{placeins}
\usepackage[T1]{fontenc}
\usepackage[utf8]{inputenc}
\usepackage{geometry}                		
\usepackage[parfill]{parskip}    		
\usepackage{graphicx}				

\usepackage[table]{xcolor}
\usepackage{adjustbox}
\usepackage{amssymb, amsmath}
\usepackage{bm} 
\usepackage[super]{natbib} 
\usepackage{tikz}
\usepackage{hyperref} 
\usepackage[capitalise]{cleveref}
\usepackage{subcaption}
\usepackage{blkarray}
\usepackage{rotating}
\usepackage{stackrel}
\usepackage{float} 
\usepackage{listings}
\usepackage{amsthm}
\usepackage[T1]{fontenc}
\usepackage[utf8]{inputenc}
\usepackage{mathtools}
\usepackage{amsmath}
\usepackage[english]{babel}
\usepackage{soul}
\usepackage{makecell}
\usepackage{caption}
\usepackage{multirow}
\usepackage{multicol}
\usepackage{authblk}
\usepackage{verbatim}
\usepackage{url}
\usepackage[symbol]{footmisc}
\usepackage{setspace}

\crefname{enumi}{Question}{Questions}

\providecommand{\keywords}[1]{\textit{Keywords:} #1}
\newcommand{\appendixnegbin}{\textrm{Section 1 of the Supplementary Materials}}
\newcommand{\appendixinlaerrors}{\textrm{Section 2 of the Supplementary Materials}}

\newlength\tindent
\usepackage{booktabs} 
\usepackage{array} 
\usepackage{paralist} 

\usepackage{fancyhdr} 
\usepackage{titlesec}
\titleformat{\section}
  {\large\bfseries}{\thesection.}{1em}{}
\titleformat{\subsection}
  {\normalsize\bfseries}{\thesubsection.}{1em}{}
  
 \titleformat{\paragraph}[runin]{\normalsize\bfseries}{\theparagraph}{1em}{}
\usepackage{lastpage}

\date{}
\definecolor{light-gray}{gray}{0.95}

\begin{document}

\title{Computing with R-INLA: Accuracy and reproducibility with implications for the analysis of COVID-19 data}

\author[1]{\small Kori Khan\footnote{Corresponding author: kkhan@iastate.edu}}
\author[2]{\small Hengrui Luo  }
\author[3]{\small Wenna Xi}
\affil[1]{\footnotesize Department of Statistics, Iowa State University}
\affil[2]{\footnotesize Lawrence Berkeley National Laboratory}
\affil[3]{\footnotesize Department of Population Health Sciences, Weill Cornell Medicine}
\date{}

\maketitle

\begin{abstract}
The statistical methods used to analyze medical data are becoming increasingly complex. Novel statistical methods increasingly rely on simulation studies to assess their validity. Such assessments typically appear in statistical or computational journals, and the methodology is later introduced to the medical community through tutorials. This can be problematic if applied researchers use the methodologies in settings that have not been evaluated. In this paper, we explore a case study of one such method that has become popular in the analysis of coronavirus disease 2019 (COVID-19) data. The integrated nested Laplace approximations (INLA), as implemented in the R-INLA package, approximates the marginal posterior distributions of target parameters that would have been obtained from a fully Bayesian analysis. We seek to answer an important question: Does existing research on the accuracy of INLA’s approximations support how researchers are currently using it to analyze COVID-19 data? We identify three limitations to work assessing INLA's accuracy: 1) inconsistent definitions of accuracy, 2) a lack of studies validating how researchers are actually using INLA, and 3) a lack of research into the reproducibility of INLA’s output. We explore the practical impact of each limitation with simulation studies based on models and data used in COVID-19 research. Our results suggest existing methods of assessing the accuracy of the INLA technique may not support how COVID-19 researchers are using it. Guided in part by our results, we offer a proposed set of minimum guidelines for researchers using statistical methodologies primarily validated through simulation studies.
\end{abstract}


\section{Introduction}
 The coronavirus disease 2019 (COVID-19) pandemic has underscored the importance of reliable and reproducible analyses in medical research. As new variants of the virus emerge, it is important that changes in the results of statistical analyses are due to changes in the virus. In medical and epidemiological research, COVID-19 data are often analyzed with disease mapping models. Historically, a fully Bayesian approach has been one of the most popular methods of analysis for such models, particularly when the data has spatio-temporal features \citep{best2005comparison,wakefield2000bayesian,wakefield2007disease,banerjee2014hierarchical}. However, employing a Bayesian analysis in this context can be a computationally expensive and time-intensive effort \citep{knorr2002block,rue2009approximate}. The conditions of the pandemic can change rapidly, so there is considerable pressure on researchers to favor faster methods of analysis.  
    	
A review of the literature analyzing COVID-19 data indicates that one of the most popular alternatives to a fully Bayesian approach involves the use of the R-INLA package. Among other things, the R-INLA package has been used in efforts to identify risk factors associated with COVID-19 \citep{dimaggio2020black,millett2020assessing}, find factors predictive of death from COVID-19 \citep{rodriguez2020risk,saez2020effects}, assess the effectiveness of preventative policies \citep{saez2020effectiveness}, and detect clusters of COVID-19 cases \citep{neyens2020can}. Statistically, these efforts have relied on R-INLA's implementation of  a rich variety of models. These  include, but are not limited to, zero-inflated negative binomial models, Poisson regression models, and binomial regression models. Furthermore, these models have included those both with and without spatial and temporal random effects \citep{dimaggio2020black,millett2020assessing,rodriguez2020risk,neyens2020can,jalilian2021hierarchical,briz2020spatio}.

The R-INLA package implements the Integrated Nested Laplace Approximation (INLA) method, which was introduced by \citet{rue2009approximate} and further refined by \citet{martins2013bayesian} (see also \citet{lindgren2011explicit}).
Analyses conducted with the R-INLA package are often referred to as Bayesian approaches in the epidemiological and medical literature. Technically, however, the INLA method is an approximation of a Bayesian approach. More specifically, for a class of latent Gaussian models, INLA is designed to approximate the marginal posterior distributions that would have been obtained from a Bayesian analysis employing Markov chain Monte Carlo (MCMC) methods \citep{ferkingstad2015improving}. 
R-INLA’s explosion in popularity predates the pandemic (for a non-exhaustive list of its use in a variety of applications, see pages 396-397 in \citet{rue2017bayesian}). Anyone who has used both MCMC and R-INLA to analyze data can understand the appeal of R-INLA: It is very fast. Until recently, research using the R-INLA primarily appeared in journals focusing on computing or new statistical methodologies. Such work, while very valuable, is typically a step away from directly shaping policy decisions. The growing popularity of the R-INLA package in COVID-19 research is a shift to a setting where the results can directly shape the medical communities policy decisions. As new variants of the virus are emerging, a natural question emerges: Are the analyses conducted with R-INLA providing us with accurate and reproducible results?
    	
In this paper, we make a first attempt to explore this question. We focus on the INLA method \textit{as implemented by the R-INLA package}, and we use the terms interchangeably unless it would cause confusion. We allow the work done on COVID-19 to shape our approach. The rest of this paper is structured as follows: In \cref{sec:INLA}, we provide a brief introduction to the INLA method as well as a clarification of our use of the term ``reproducible.'' We point out that there are few theoretical guarantees about the accuracy of INLA's approximations, and we review existing efforts to assess this accuracy. We identify three key limitations to the scope of these efforts in settings, like COVID-19 research, where analyses are used to shape public policy. In \cref{sec:simulation}, we design a series of simulations, informed by the types of models and data appearing in COVID-19 literature, to further explore these three limitations. In \cref{sec:casestudies}, we select a peer-reviewed article whose results have been referenced both in the academic literature and in American mainstream media \citep{millett2020assessing,barron2020new}. We use the actual data and, when possible, the INLA code employed in \citet{millett2020assessing} to try to reproduce the results that were previously obtained. We also explore how well the results obtained by INLA approximate the corresponding results that would have been obtained if the researchers had instead used MCMC to fit the model. Finally, in \cref{sec:discussion}, we draw on the insights we gained through our explorations. We offer a set of minimum practical guidelines for medical researchers future use of statistical methodologies, like INLA, which have been primarily validated through simulation studies.

\section{Background} 

In this section, we provide a short summary of background material the rest of the paper relies on. We emphasize that this section, and the rest of the paper, are motivated by analyses of COVID-19 data which are used to shape public policy. Intuitively, it is uncontroversial that we, as a society, want the analyses of COVID-19 to be dependable enough that we can understand what is driving the spread of a novel disease as well as how this story changes with new variants. There is a long history of analyses of human disease playing a role in shaping public policy, and an understanding that such analyses need to be trustworthy  \citep{peng2006reproducible,samet2000epidemiology}. 

With this motivation in mind, in \cref{sec:INLA}, we provide a brief overview of the INLA method and how its output has been assessed in peer-reviewed literature. In the scientific community, ``dependability'' and ``trustworthines'' are typically mapped to concepts of reproducibility and replicability. In this paper, we will focus on reproducibility, and in \cref{reproduce}, we take a moment to clarify our use of the term ``reproducible.''

\subsection{A Brief Overview of INLA} \label{sec:INLA}

Integrated nested Laplace approximation (INLA) is a recently proposed method for approximate Bayesian inference for structured additive regression models with latent Gaussian fields \citep{rue2009approximate,martino2009implementing,ferkingstad2015improving}. In this subsection, we provide a brief overview of the general approach in order to highlight sources of potential error in the approximations INLA achieves.  We refer interested readers to \citet{rue2009approximate}, \citet{martins2013bayesian}, and \citet{ferkingstad2015improving} for more details. We acknowledge that various extensions and variations to the INLA approach we will describe here have been proposed \citep{lindgren2015bayesian}. However, we focus on only the general approach detailed in \citet{rue2009approximate} and \citet{martins2013bayesian} and explicitly exclude the stochastic partial differential equations (SPDE) approach..

The literature on INLA typically separates out model parameters into two groups: those that are given Gaussian priors (often denoted $\bm{x}$) and those that are not given Gaussian priors (often denoted $\bm{\theta}$). The latter are called hyperparameters in INLA related literature. We adopt this terminology for this paper, while acknowledging that in some cases it does not agree with how the term is used in the broader Bayesian literature. The goal of INLA is to approximate, for all $x_i$:
\begin{equation}\label{marginal}
    \pi(x_i | \bm{y} ) = \int \pi(x_i | \bm{\theta}, \bm{y}) \pi (\bm{\theta} | \bm{y} )~ d \bm{\theta}. 
\end{equation} 
Here, $\bm{y}$ represents the observed data with likelihood $\pi (\bm{y} |  ~ \bm{x} )$. INLA proposes to accomplish this approximation in two stages. 

In the first step of INLA, the marginal posterior density of $\pi (\bm{\theta} | \bm{y})$ is approximated with a Laplace-like approximation \citep{martins2013bayesian}. The purpose of this step is to numerically integrate out $\pi( \bm{\theta} | \bm{y})$ in Equation \eqref{marginal}.  The result of this step is a set of weighted points for the hyperparameters which can be used to approximate $\pi (x_i | \bm{\theta}, \bm{y})$.
As for the second step, the current implementation of INLA allows for three ways to accomplish the approximation:  the Gaussian, the simplified Laplace, and the Laplace approximation \citep[see][]{rue2009approximate}. We have listed these options in an increasing order of both computational expense and supposed accuracy.

To our knowledge, the only theoretical guarantees of INLA's accuracy in peer-reviewed literature are related to the asymptotic validity of the Laplace approximation itself (outside the context of INLA) \citep{rue2009approximate}. However, INLA is used in many settings where the conditions that ensure the validity of the Laplace approximation are \textbf{not} met \citep{rue2009approximate,fong2010bayesian,ferkingstad2015improving}. Additionally, there are sources of error beyond the error related to the Laplace approximation. A full discussion of these additional sources of error is beyond the scope of this paper, but we offer an intuitive overview of some of them in \appendixinlaerrors ~ ( see also the responses to \citet{rue2009approximate}). Some of the factors that can influence the accuracy of INLA include the likelihood  $\pi( \bm{y} | \bm{\theta})$, the priors chosen for $\bm{\theta}$, and the data itself. 

The absence of theoretical guarantees for the accuracy of INLA's approximations is not, in itself, problematic. However, if we again consider the implications of using INLA in analyses which in turn shape public policies, it is important that the accuracy of its approximations be carefully reviewed. To do so, we turn to the current literature assessing INLA. Despite the fact that the accuracy of INLA's approximation will likely depend on the model used, claims about the accuracy of INLA's approximations are typically broad. It is not uncommon to find statements in the literature stating that INLA's approximations are ``accurate'' or ``extremely accurate'' for a wide variety of applications. \citep{rue2009approximate,schrodle2011spatio,martins2013bayesian,lindgren2011explicit}. A closer look reveals that some of the papers that include statements like these do not provide an assessment of the accuracy of INLA's output. \citep{lindgren2015bayesian,blangiardo2013spatial,schrodle2011spatio}. Those papers that do assess the accuracy of INLA tend to consider one of two categories of accuracy measures.

In the first category, accuracy is defined by how close the output from INLA is to the output from MCMC. Typically, a model is fit with both MCMC and INLA to a single dataset (or more rarely, a series of simulated datasets).  The results of the two methods of analysis are then compared \citep{rue2009approximate,fong2010bayesian,held2010posterior,ferkingstad2015improving}. The most common method of comparing the output of INLA to the output of MCMC is the use of plots. In particular, a histogram of the sample obtained from a parameter's marginal posterior distribution from the MCMC method is  overlaid with a density plot of the corresponding marginal posterior distribution obtained from the INLA method \citep{rue2009approximate,ferkingstad2015improving}. Another common method is to compute how close the marginal posterior means and variances from INLA are to the corresponding estimates from MCMC. In the second category, accuracy is related to  whether the estimates from INLA or MCMC are closer to the ``truth'' (the value used to generate  data in a simulation study) \citep{carroll2015comparing,ferkingstad2015improving,taylor2014inla}. Often, if the results from MCMC are not clearly ``better'' than the results from INLA, INLA is thought to have performed well.

There are several important limitations to the existing work assessing the accuracy of INLA's output. In this paper, we focus on three specific limitations we feel are most relevant to researchers wishing to use INLA for work that will inform public policy. The first limitation stems from the inconsistent measures of accuracy. To our knowledge, no one has simultaneously compared the output from INLA to the output from MCMC \textit{and} to the generating value in a simulation study. In other words, we do not know whether the cases in which INLA failed to approximate the results from MCMC well are the same cases in which INLA failed to approximate the generating value well. The ambiguity in these definitions could mean that the same output from INLA could lead to inconsistent conclusions regarding accuracy. From a practical standpoint, this is problematic because both types of measures are currently used to support the use of INLA.

A second limitation involves inconsistencies between how INLA is assessed and how practitioners are using INLA. In practice, many of the inferential methods which statisticians and applied researchers use rely on the tails of the distribution (e.g., a 95\% credible interval). Many of the summaries of INLA's accuracy, in particular the ones that depend on the plots, will give little indication of how well INLA approximates the  characteristics of such tail behavior. More broadly, researchers typically do not rely on marginal posterior distributions alone for inference. For example, researchers commonly use functions of parameters (requiring the summary of a transformation of the marginal posterior distributions) or various model selection criteria. With a few exceptions, such as \citet{held2010posterior}, INLA's ability to calculate these transformations or criteria are often not assessed in peer-reviewed literature.

Finally, there is a third and more subtle limitation to the current methods of assessing INLA's accuracy -- one we have yet to see referenced anywhere else. To our knowledge, no one has ever assessed whether the analyses conducted by INLA, as implemented in the R-INLA package, are reproducible (in the sense specified in \cref{reproduce} below). Since the accuracy of the INLA method has been primarily supported by illustrative examples and simulation studies, we believe that illustrative examples and simulation studies should be, at the minimum, reproducible. Reproducibility, however, has many definitions. In the next subsection, we clarify our use of the word in this paper.

\subsection{Reproducibility} \label{reproduce}

As we have already referenced, scientists typically think of ``trustworthiness'' in terms of either reproducibility or replicability. There are various definitions of reproducibility and replicability in use, varying depending on the field \citep{kenett2015clarifying,goodman2016does,plesser2018reproducibility,peng2021reproducible}. Most typically, reproducibility has to do with obtaining the same results with the same analysis on the same data and replicability has to do with obtaining the same (or similar) results with new data \citep{peng2021reproducible}.

With respect to the current use of INLA, we found that issues related to reproducibility rather than reliability are most relevant. In fields where analyses are used to shape public health policies, researchers have argued that a minimum level of reproducibility should involve, among other things, the ability to reproduce an analysis on a given dataset with the same or ``similar'' methods \citep{schwab2000making,peng2006reproducible,peng2011reproducible}. In this paper, we consider a version of reproducibility weaker than this: Given the same data and the same code, can the same results be obtained? With this clarification in mind, we now proceed to use simulation studies to explore the three limitations identified in \cref{sec:INLA}.

\section{Simulation Studies} \label{sec:simulation}
In this section, we designed a set of simulation studies to explore each of the three limitations identified in \cref{sec:INLA} in the context of the analysis of COVID-19 data. For convenience, we now express these limitations as questions:

\begin{enumerate}
    \item \label{quesone} Do the two common concepts of accuracy lead to consistent conclusions about INLA's approximations? 
    \item \label{questwo} Do the current methods of assessing the accuracy of INLA's approximations support how practioners are using INLA in the analysis of COVID-19 data? 
     \item \label{questhree} Are results obtained with INLA reproducible?
\end{enumerate}

As an overview, in \cref{sim:poisson,sim:bym}, we focus on \cref{quesone}. We also share, when relevant, insights we gained about \cref{questwo}. In \cref{sim:model selection}, we focus on \cref{questwo}. While we had hoped to thoroughly investigate \cref{questhree} in the context of these simulation studies, we found there were simply too many variables to control for. A myriad factors, including the data analyzed, the statistical model used, the version of R-INLA used, and the operating system the code was run on impacted reproducibility. We offer our observations about the implications these various factors had on the results for each of the simulation studies. We then take these observations and offer a more thorough exploration of \cref{questhree} in the context of a single dataset in \cref{sec:casestudies}.

In designing simulation studies for this section, we kept in mind both how INLA has previously been evaluated and how it is currently being used in practice to analyze COVID-19 data. To ensure that the simulation studies mirrored the ways that researchers are using INLA to analyze COVID-19 data, we first reviewed the literature analyzing COVID-19 data. Much of it involves the analysis of count data both with and without spatial random effects \citep{dimaggio2020black,lima2021spatial,neyens2020can, briz2020spatio,konstantinoudis2020long,rodriguez2020risk,millett2020assessing}. Thus, in the following simulation studies, we generated count data from models both with and without spatial random effects. We then used INLA and MCMC methods to fit two different models to this generated data. Following the trends in the literature, both of these models rely on the Poisson distribution, as described in more detail in \cref{sim:poisson,sim:bym}.

To explore \cref{quesone}, we first defined two measures of accuracy. To evaluate how well the INLA output approximated the output from MCMC for an arbitrary parameter $\theta$, we used an adaption of the measure put forth by \citet{fong2010bayesian}. We call this measure ``Percent Error'' (PE) and define it as:
\begin{equation} \label{eq:error}
 \textrm{PE} \left( \theta \right) = \left( \frac{ \textrm{E}[ \theta_{\textrm{INLA}} | \bm{Y} ] - \textrm{E} [ \theta_{\textrm{MCMC}} | \bm{Y} ]  }{\textrm{SD}[\theta_{\textrm{MCMC}} | \bm{Y} ]} \right) \times 100,
\end{equation}
where $\textrm{E} [ \theta_{\textrm{INLA}} | \bm{Y}]$ is the marginal posterior mean for $\theta$ from the INLA output, $\textrm{E} [ \theta_{\textrm{MCMC}} | \bm{Y}]$ is the marginal posterior mean for $\theta$ from the MCMC output, and $\textrm{SD}[\theta_{\textrm{MCMC}} | \bm{Y} ]$ is the marginal posterior standard deviation for $\theta$ from the MCMC output.  Intuitively, \cref{eq:error} measures the size of the error between INLA and MCMC as a percentage of the ``true'' error. In the supplemental material of \citet{fong2010bayesian}, the authors considered absolute values of around 30\% as potentially problematic, and absolute values of 20\% or under as acceptable. We adopted these rough guidelines for discussion, with the understanding that different researchers might have different tolerances for relative errors.

To evaluate how close the MCMC and INLA estimates are to the generating value, we consider the percent change (PC) defined in \cref{eq:pc}. We choose this measure because, unlike other commonly used statistics, percent change has the same scale for different parameters. To calculate the percent change for an arbitrary parameter $\theta$, we used the following formula:
\begin{equation} \label{eq:pc}
\textrm{PC}_{\textrm{Method}}\left(\theta \right) = \left( \frac{ \textrm{E} \left[ \theta_{\textrm{Method}} | \bm{Y} \right] - \theta_{\textrm{GV}}}{\theta_{\textrm{GV}}} \right) \times 100, 
\end{equation}
where $\textrm{E} [ \theta_{\textrm{Method}} | \bm{Y}]$,  $\textrm{Method} \in \{\textrm{MCMC}, \textrm{INLA}\}$, is the marginal posterior mean for $\theta$ from either the MCMC or INLA output, and $\theta_{\textrm{GV}}$ is the generating value used to simulate the data.

To explore \cref{questwo}, we again reviewed how researchers were using INLA in the COVID-19 literature. We found that many were using  model selection criteria available in the R-INLA package, including the conditional predictive ordinate (CPO), Watanabe-Akaike information criterion (WAIC), and deviance information criterion (DIC) \citep{saez2020effectiveness,jalilian2021hierarchical,briz2020spatio}. These works typically involved using the model selection criterion to choose between models with and without the inclusion of temporal or spatial random effects. Thus, we designed our simulation studies so that we could evaluate INLA's calculation of WAIC for model selection between a model without a spatial random effect and a model with a spatial random effect. 

Before delving into the details of the simulation studies, we note it is important to ensure that the priors and distributions used in the MCMC approach match those used in the R-INLA package. Therefore, we used the R package NIMBLE \citep{nimble-software:2021} for the MCMC approach analyses, which allows for user-defined distributions and priors \citep{nimble-manual:2021,nimble-article:2017}. In the INLA literature, running MCMC chains for $1,000,000$ iterations and discarding $100,000$ is sometimes referred to as the ``gold-standard'' for comparison to INLA \citep{ferkingstad2015improving}. To be careful, for all fitted models in \cref{sim:poisson,sim:bym,sim:model selection}, the MCMC chains were run for $2,000,000$ iterations with a burn-in of $100,000$. When fitting models with the R-INLA package, we used stable versions of the package. We also used the Laplace approximation for all models because it is considered the most accurate. All code and data used in this section are available as described in \cref{data_code}.

\subsection{Poisson Regression Model} \label{sim:poisson}

We began our simulation study by generating 100 datasets from a Poisson regression model. To ensure that the generated data mirrored real-world COVID-19 data, we relied on actual data for each of the $296$ census tracts located in Milwaukee County, Wisconsin, and a model inspired by the one used by \citet{dimaggio2020black} to analyze COVID-19 data in New York City.

For each of the 100 datasets, the response variable $y_i$, $i = 1, \ldots, 296$, was generated conditionally independently from a Poisson distribution with rate parameter $\lambda_i$: 
\begin{equation*}
\log \left( \lambda_i \right) = .1 + .05 x_{i} +  \log\left(\textrm{Total}_i \right)  + \epsilon_i,
\end{equation*} 
where $x_{i}$ is the percent of individuals whose income is below the poverty rate in the $i$th census tract, $\textrm{Total}_i$ is the total number of COVID-19 tests recorded between August 1, 2020 and October 15, 2020 in the $i$th census tract, and $\epsilon_i$ is the random error for the $i$th census tract, $i = 1, \ldots, 296$. Covariates $x_{i}$ and $\textrm{Total}_i$ came from actual data for census tracts in Milwaukee County (see \cref{data_code} for details on the data), and $\epsilon_i$ was generated identically independently from $\mbox{Normal}(0, 1)$. Covariates $x_{i}$ and $\textrm{Total}_i$ remained fixed for the generation of each dataset.

For each simulated dataset, we then fit the following Poisson regression model with both INLA and MCMC methods:
\begin{equation*}
    y_i | \lambda_i \stackrel{\mbox{ind}}{\sim} \textrm{Poisson}(\lambda_i),
\end{equation*}
where $\lambda_i$ was modeled as:
\begin{equation*}
    \log(\lambda_i)  =  \beta_0 + \beta_x x_i + \log(\textrm{Total}_i) + \epsilon_i,
\end{equation*}
with $\epsilon_i \stackrel{\mbox{iid}}{\sim} \mbox{Normal}(0, \sigma)$. Here, $\sigma$ is a precision parameter. We used the default priors described in the documentation of the R-INLA software: both $\beta_0 $ and $\beta_x$ were given normal priors with mean $0$ and standard deviation $1000$ and $\textrm{log}(\sigma)$ was given a log-gamma prior with shape parameter 1 and scale parameter .00005.

\begin{figure}[h] 
	\centering
	\subcaptionbox{}{\includegraphics[width=0.45\textwidth]{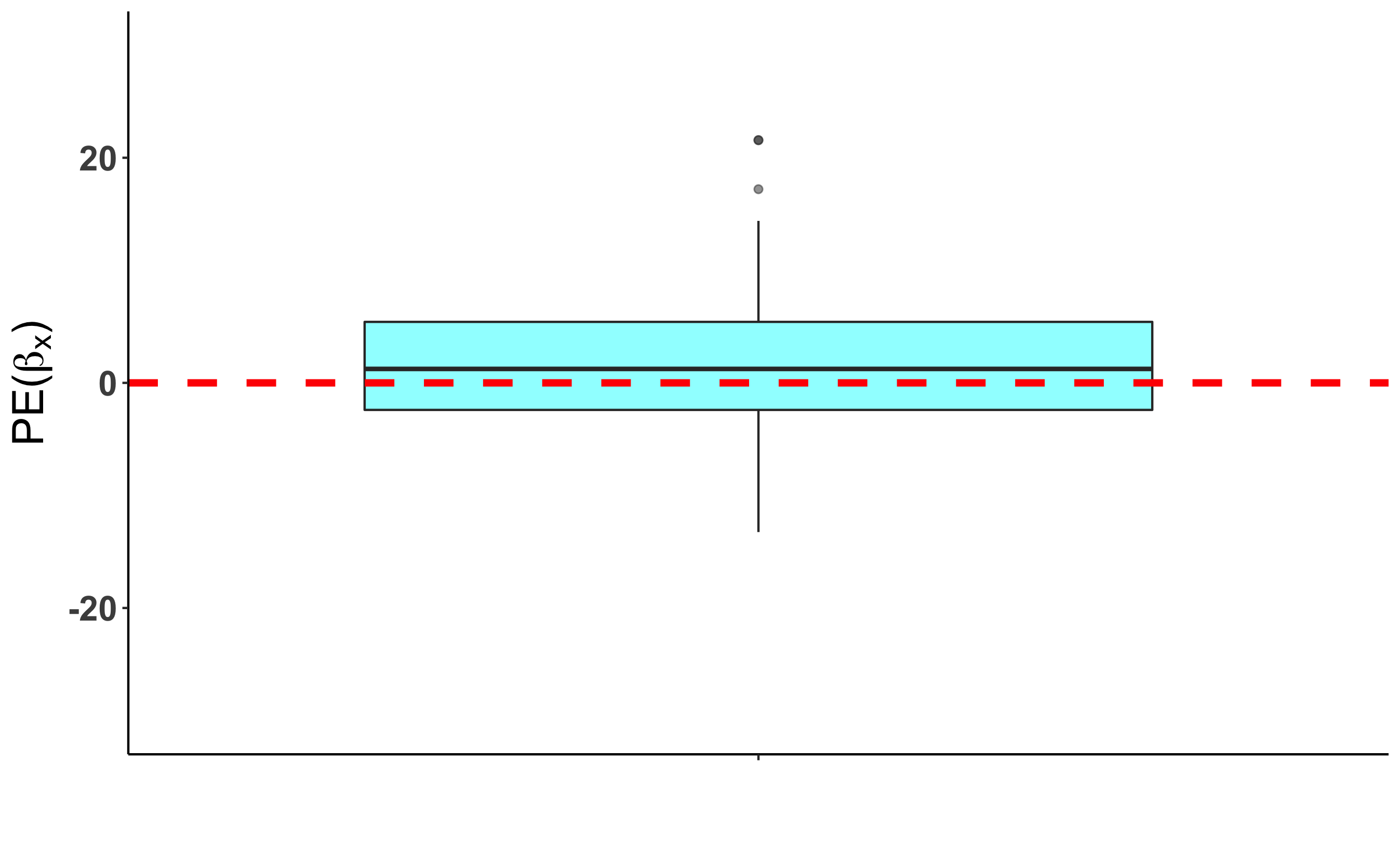}}
	\hfill 
	\subcaptionbox{}{\includegraphics[width=0.45\textwidth]{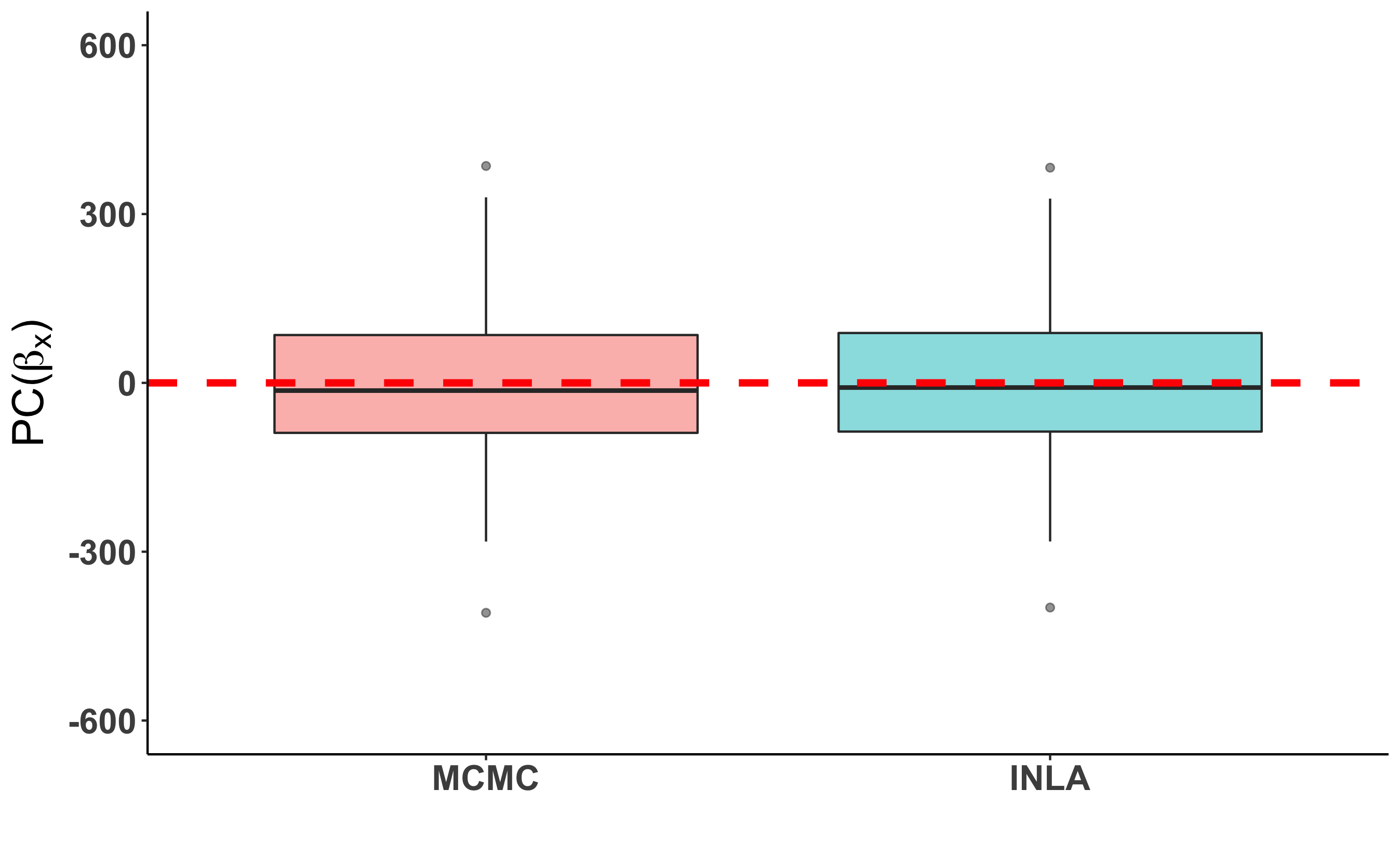}}%
	\caption[Description]{Boxplots of comparing posterior estimates of $\beta_x$ for nonspatial data. (a) Percent error (PE). (b) Percent changes (PCs) from MCMC and INLA to the generating value 0.05.}
	\label{fig:poisson_beta}
\end{figure}

\begin{figure}[h] 
	\centering
	\subcaptionbox{}{\includegraphics[width=0.45\textwidth]{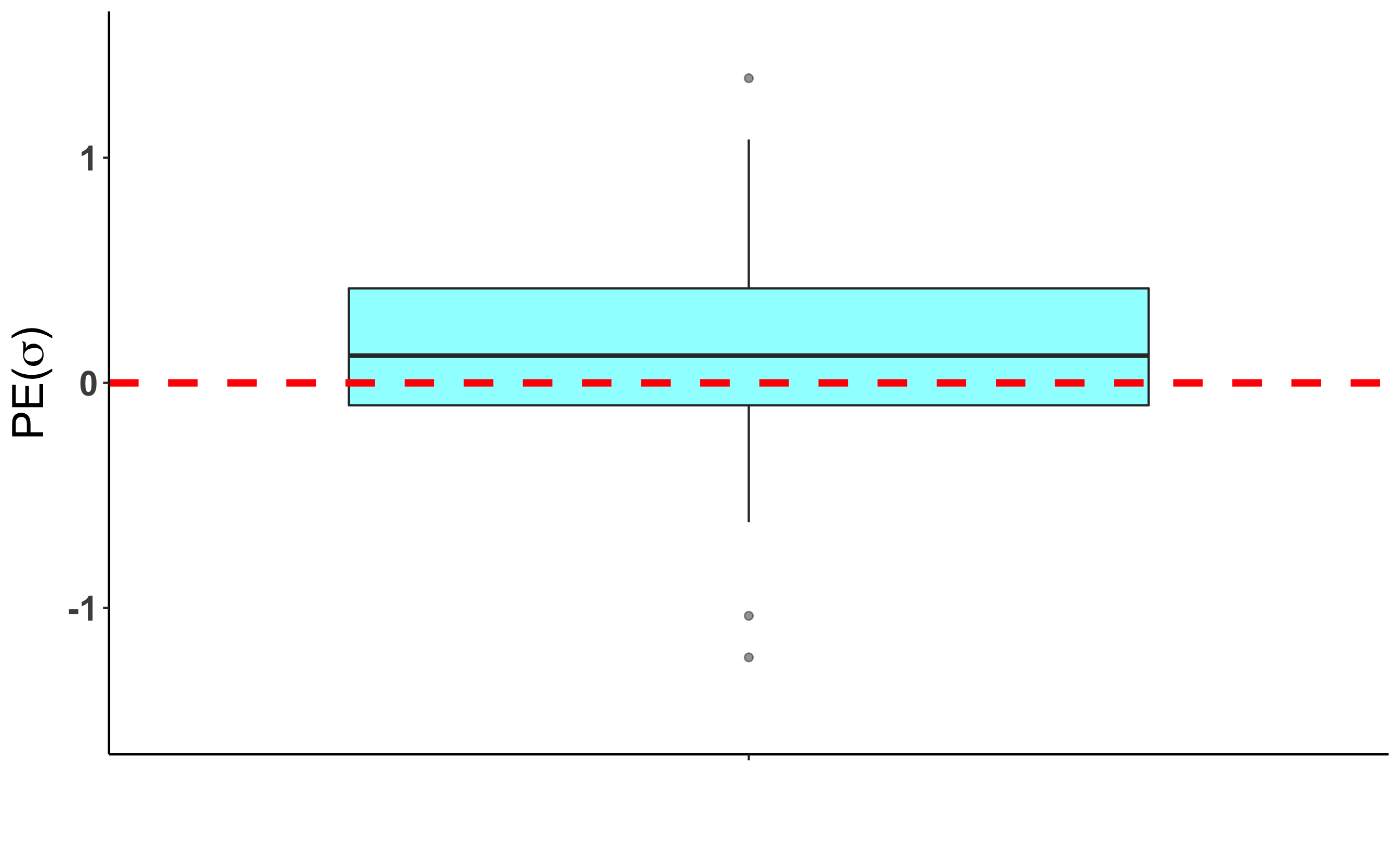}}%
	\hfill 
	\subcaptionbox{}{\includegraphics[width=0.45\textwidth]{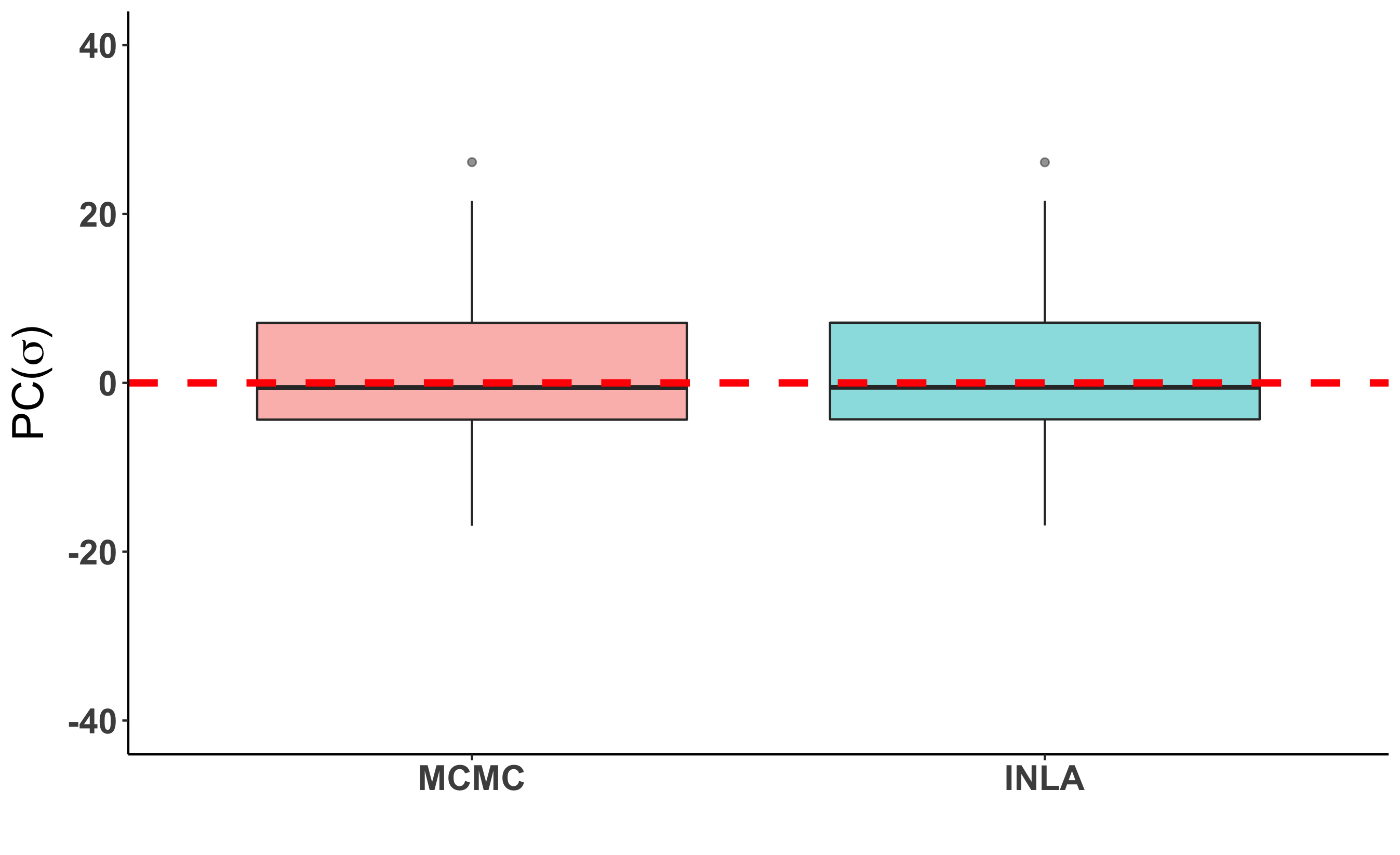}}
	%
	
	\caption[Description]{Boxplots of comparing posterior estimates of $\sigma$ for non-spatial data. (a) Percent error (PE). (b) Percent changes (PCs) from MCMC and INLA to the generating value 1.}
	\label{fig:poisson_sigma}
\end{figure}

We focused on the parameters $\beta_x$ and $\sigma$ in the following analyses. For each dataset, we first calculated $\textrm{PE}(\beta_x)$ and $\textrm{PE}(\sigma)$, as defined in \cref{eq:error}, to compare the posterior estimates from INLA to the posterior estimates from MCMC. \cref{fig:poisson_beta}(a) is the boxplot summary of the 100 $\textrm{PE}(\beta_x)$'s, and \cref{fig:poisson_sigma}(a) is the boxplot summary of the 100 $\textrm{PE}(\sigma)$'s. For both parameters, we observe that the results from INLA and MCMC were quite similar. The $\textrm{PE}(\beta_x)$'s were almost always less than 20\%. The results between INLA and MCMC matched even more closely for estimates of $\sigma$: The $\textrm{PE}(\sigma)$'s were almost always less than 1\%. We then used \cref{eq:pc} to compare the posterior estimates from MCMC and INLA to the generating value (.05 for $\beta_x$ and 1 for $\sigma$). The results for $\beta_x$ are summarized in \cref{fig:poisson_beta}(b) and the results for $\sigma$ are summarized in \cref{fig:poisson_sigma}(b). Both MCMC and INLA led to better estimates for $\sigma$ than they did for $\beta_x$. For both $\beta_x$ and $\sigma$, the distributions of PC from the generating value looked quite similar for MCMC and INLA. 

We consider what these results mean with respect to \cref{quesone}. As currently used in the literature, the two kinds of accuracy represented by our $\textrm{PE}$ and $\textrm{PC}$ calculations support the accuracy of INLA. INLA seemed to well approximate the results from MCMC for both $\beta_x$ and $\sigma$. Furthermore, INLA and MCMC appeared to do about the same in estimating the generating value. However, with respect to $\beta_x$, this approximation was not very good. 

We now take a moment to explain our exploration of \cref{questhree} in this simulation study. We began by attempting to run an analysis of a single dataset on the same operating system with the same version of R-INLA. We found that the summary of the posterior marginal distributions for $\beta_x$ and $\sigma$ could change, although not significantly. We were able to rectify this by setting the number of threads (\texttt{num.threads} argument) the INLA program could use to 1.
We used four combinations of the two operating systems and the two INLA versions described in \cref{data_code}. We found that both the version of R-INLA used and the operating system could impact the results obtained. For thoroughness, we also assessed whether two different operating systems and two different versions of the NIMBLE package could impact the results of an MCMC analysis for one dataset. We found that there were no differences (up to tolerance .001) in the results we obtained when the seed was set. 

In this section, the results we reported and the images we created were based on the outcomes from running the simulation study once (for the operating system and R-INLA version, see \cref{data_code}). However, given our initial experience with INLA, we did rerun the simulation studies discussed in this section for INLA with the other three combinations of operating system and package version. In our experience, the version of R-INLA was associated with larger changes to results than the operating system for this simulation study. The overall patterns of accuracy measures we considered never really changed, however, and the summaries of the posterior distributions across all 100 datasets for $\beta_x$ and $\sigma$ (including the mean, 2.5\% percentile, 97,5\% percentile, and standard deviation) were within .05 of each other.

Finally, we would like to point out some observations relevant to \cref{questwo}. First, the generating value for $\beta_x$ (.05) was based on the values obtained in the analyses considered in \citet{dimaggio2020black}, and so are relevant to the magnitude of regression coefficients considered in the literature. However, the default priors used in INLA did not lead to particularly good estimates of $\beta_x$ (we found that changing the prior on $\sigma$ could improve the estimates considerably). In practice, when we were able to review the code used in papers analyzing COVID-19 data with INLA, no one ever changed or discussed the default priors used in INLA \citep{millett2020assessing,dimaggio2020black}.

\subsection{Besag-York-Molli\'e (BYM) Model} \label{sim:bym}

As we have previously noted, many of the analyses on COVID-19 data included spatial random effects. In this section, we generate data from a model often referred to as the Besag-York-Molli\'e (BYM) model \citep{besag1991bayesian}. The BYM model is a hierarchical Bayesian Poisson model which simultaneously accounts for both spatial and non-spatial heterogeneity. The spatial random effect used in the BYM model has been used in quite a few analyses of COVID-19 data \citep{dimaggio2020black,neyens2020can,briz2020spatio}. Intuitively, the spatial random effect accounts for spatial clustering in data by introducing an underlying graph to areal data, where the areal units are the vertices and edges exist between neighboring areas \citep{besag1991bayesian}.

More specifically, we generated 100  datasets for the 296 census tracts of Milwaukee County, Wisconsin, introduced in \cref{sim:poisson}. For each dataset, $y_i$ was generated conditionally independently from a Poisson distribution with rate parameter $\lambda_i$:
\[ \log( \lambda_i ) = .1 + .05 X_{1i} +  \log\left(\textrm{Total}_i \right)  + \mu_i + \epsilon_i,\]
where $x_{i}$ and $\textrm{Total}_i$ are the same fixed effects as defined in \cref{sim:poisson}, $\epsilon_i$ is the non-spatial random effect generated identically and independently from a standard normal distribution for $i = 1, \ldots, 296$, and $\bm{\mu} = \left( \mu_1, \ldots, \mu_{296} \right)$ is the spatial random effect, which is generated from the intrinsic conditional autoregressive (ICAR) prior used in the BYM model. The ICAR prior is not a proper probability distribution, and is defined as:
\begin{eqnarray} \label{icar}
p(\bm{\mu}| \tau) \propto \tau^{n-k}  \exp \left\{ \frac{-\tau}{2} {\bm{\mu}}^T \bm{Q} \bm{\mu} \right\}.
\end{eqnarray}
Here, $\tau$ is a precision parameter, $\bm{Q}$ is the graph Laplacian of the underlying graph alluded to in the previous paragraph, and $k$ is the number of connected components of that graph (for us, $k=1$). Because the ICAR prior is not proper, there are various ways to generate a realization from it. For our simulations, we generated a realization  using conditioning by kriging with the associated precision parameter $\tau = 1$ \citep{rue2005gaussian}. We note that the choice of $\sigma$ = $\tau=1$ was chosen, in part, to be the same as a previous simulation study assessing INLA's ability to estimate generating values  \citep{carroll2015comparing}. We will compare some of their results to ours at the end of this subsection.

For each simulated dataset, we then fit the BYM model with both INLA and MCMC. Due to the ICAR prior's impropriety, identifiability issues can arise if an intercept is included in the BYM model. There are two ways to handle this \citep{Pac}. The first way involves enforcing sum-to-zero constraint(s) when sampling $\bm{\mu}$. Historically, most software packages, including NIMBLE, have enforced the sum-to-zero constraint(s) with a method known as ``centering on the fly'' \citep{nimble-manual:2021,thomas2004geobugs,banerjee2014hierarchical}. The R-INLA package appears to enforce sum-to-zero constraint(s) with ``conditioning by kriging'' \citep{rue2005gaussian,goicoa2018spatio}. These two methods of enforcing sum-to-zero constraint(s) are not equivalent. The second way of handling the identifiability issue is to omit the intercept in the BYM model \textit{and} not enforce any sum-to-zero constraint(s)\citep{Pac}. 

Because we wanted to ensure that the INLA and MCMC approaches were equivalent, we chose to omit the intercept and refrain from enforcing any sum-to-zero constraints. For our simulations, this is equivalent to including an implicit intercept for both the INLA and MCMC approaches. More precisely, we fit the following model:
\begin{equation*}
    y_i | \lambda_i \stackrel{\mbox{ind}}{\sim} \textrm{Poisson}(\lambda_i),
\end{equation*}
where $\lambda_i$ was modeled as
\begin{equation*}
    \log(\lambda_i)  =  \beta_x x_i + \log(\textrm{Total}_i) + \mu_i + \epsilon_i,
\end{equation*}
with $\epsilon_i \stackrel{\mbox{iid}}{\sim} \mbox{Normal}(0, \sigma)$ and $\bm{\mu} = \left( \mu_1, \ldots, \mu_{296} \right)$  given the ICAR prior, $p(\bm{\mu}| \tau)$ described in Equation \eqref{icar}. We used the default priors described in the documentation of the R-INLA package: $\beta_x$ was given a normal prior with mean $0$ and standard deviation $1000$, $\textrm{log}(\sigma)$ and $\textrm{log}(\tau)$ were given log-gamma priors with shape parameter 1 and scale parameter .0005.

In the simulation studies, INLA failed to fit 4 of the 100 datasets. After some experimentation, we noted that rerunning the code or using the \texttt{inla.rerun} function would eventually result in the R-INLA package successfully fitting the BYM model. However, we could not ensure that the results obtained in this way were reproducible (even on the same machine). Therefore, all results involving INLA in this section only involved 96 datasets that were successfully fit without the \texttt{inla.rerun} function.

\begin{figure}[h] 
	\centering
	\subcaptionbox{}{\includegraphics[width=0.45\textwidth]{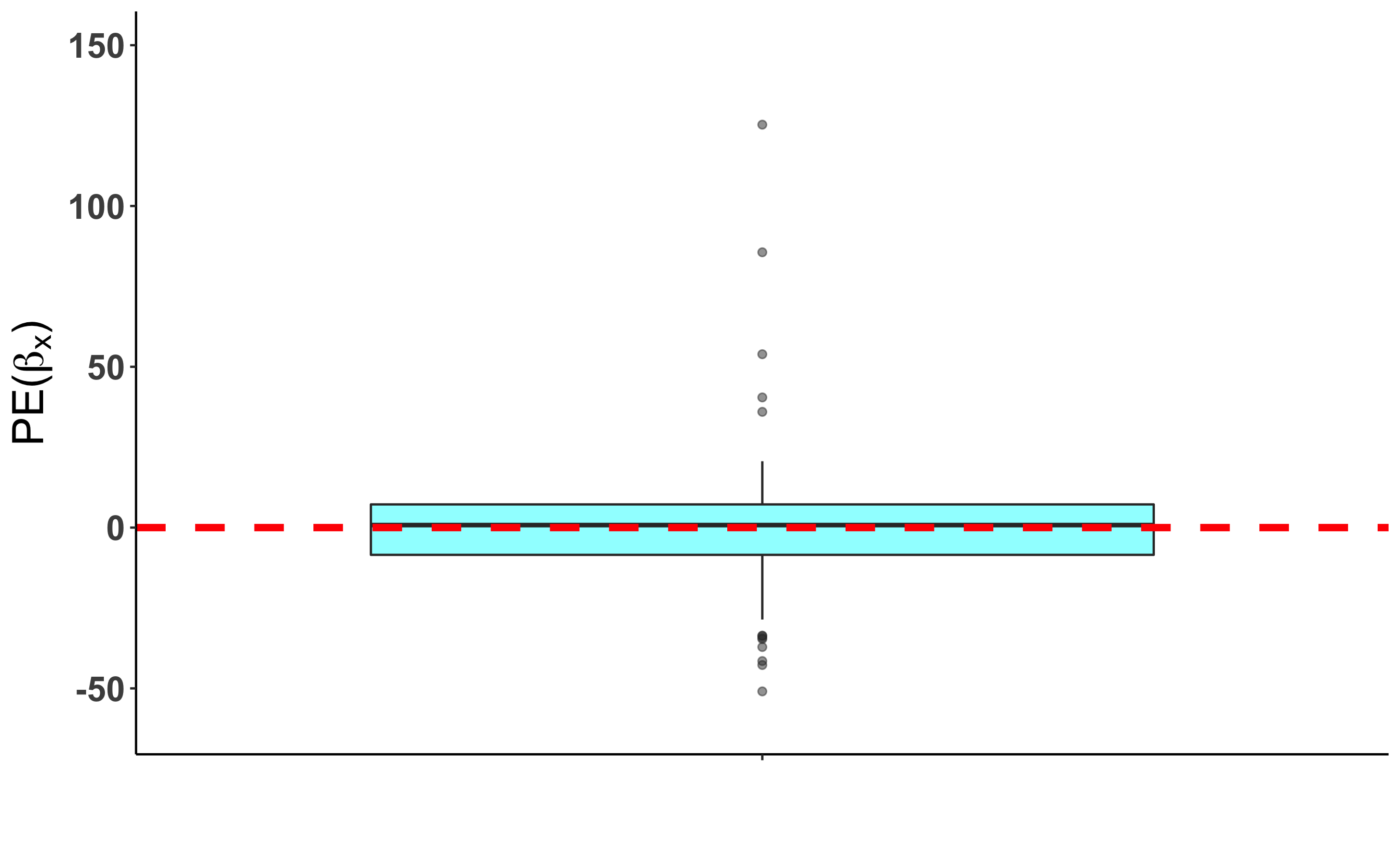}}
	\hfill 
	\subcaptionbox{}{\includegraphics[width=0.45\textwidth]{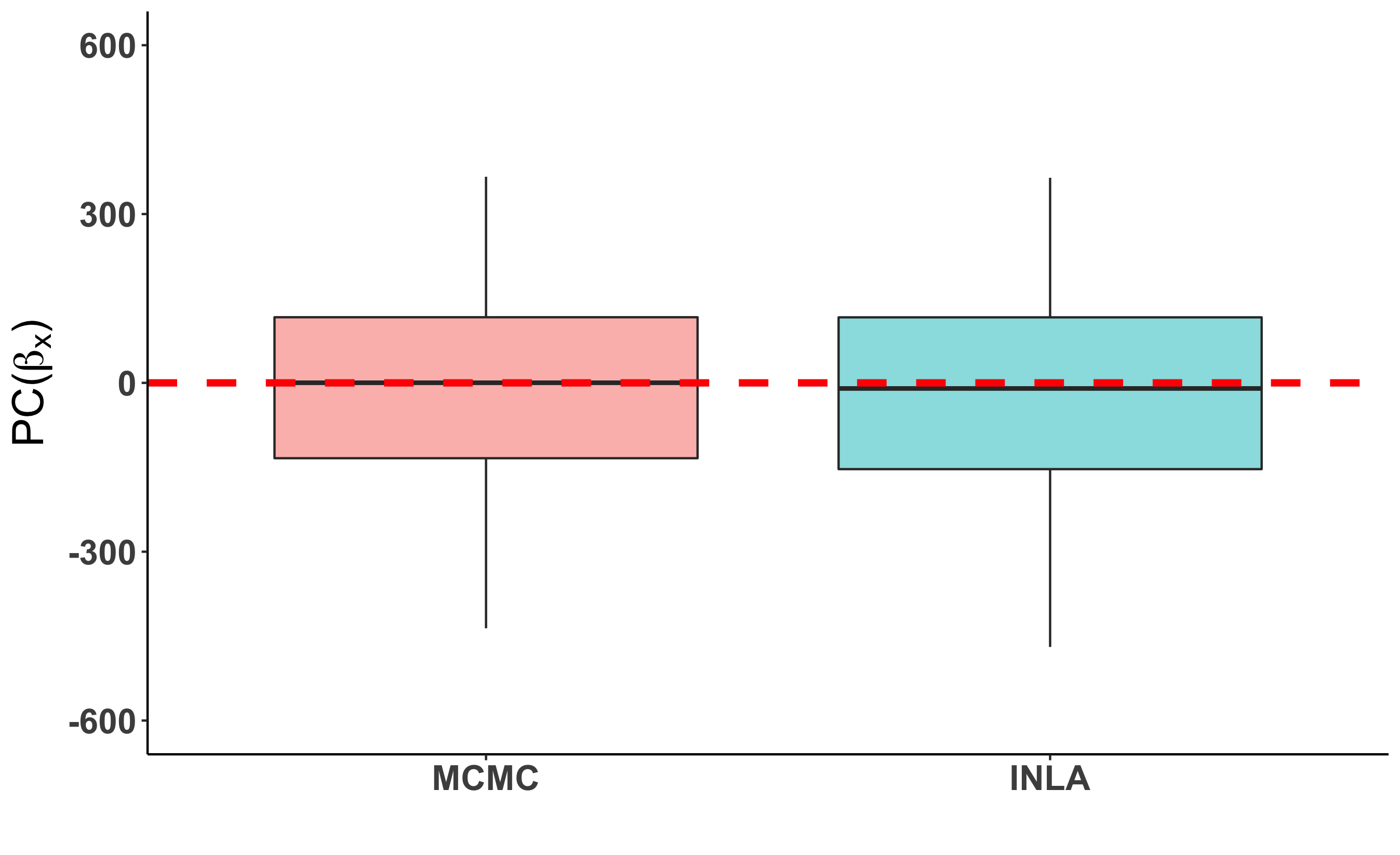}}%
	
	\caption[Description]{Boxplots of comparing posterior estimates of $\beta_x$ for spatial data. (a) Percent error (PE). (b) Percent changes (PCs) from MCMC and INLA to the generating value 0.05.}
	\label{fig:bym_beta}
\end{figure}

\begin{figure}[h] 
	\centering
	\subcaptionbox{}{\includegraphics[width=0.45\textwidth]{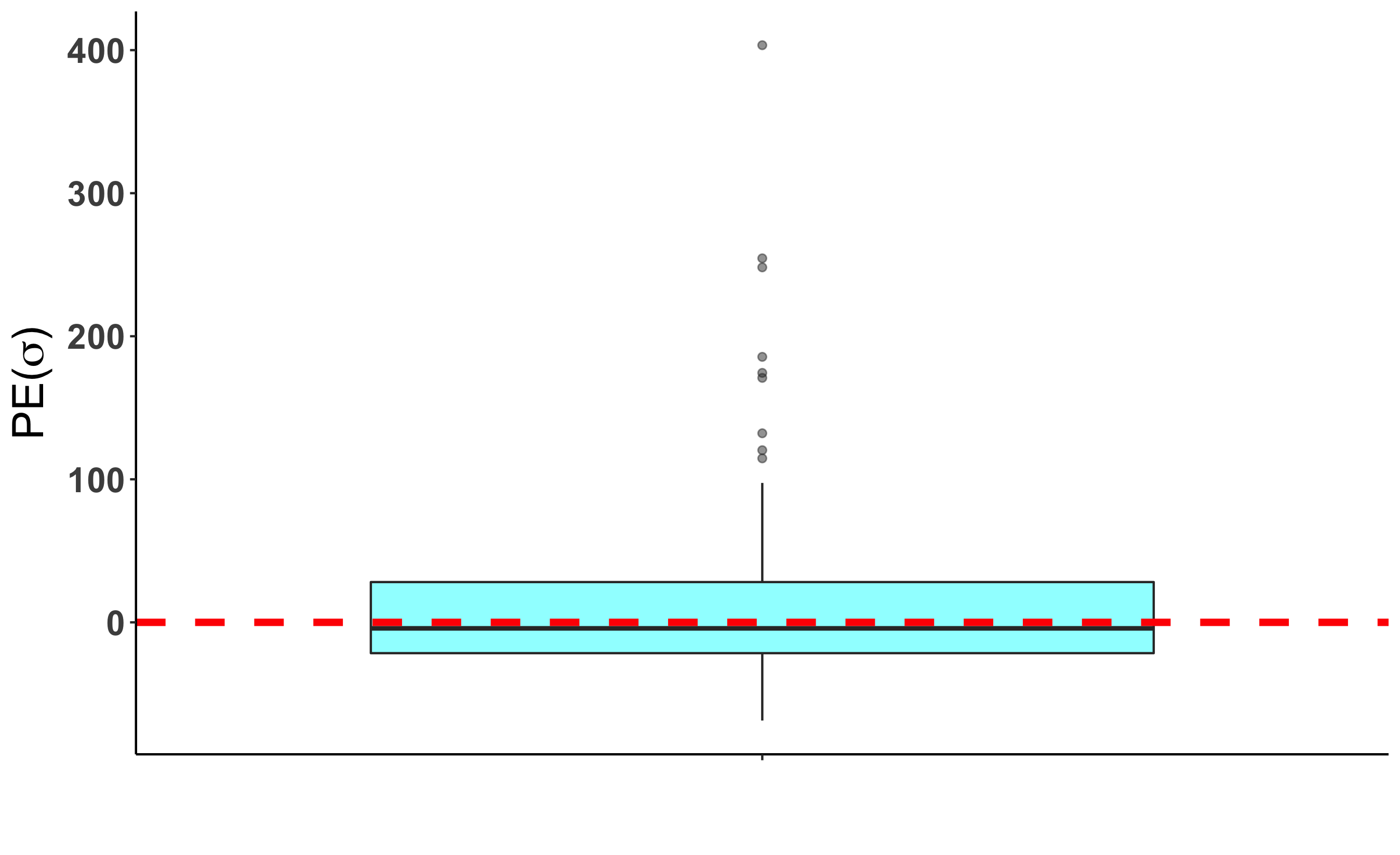}}%
	\hfill 
	\subcaptionbox{}{\includegraphics[width=0.45\textwidth]{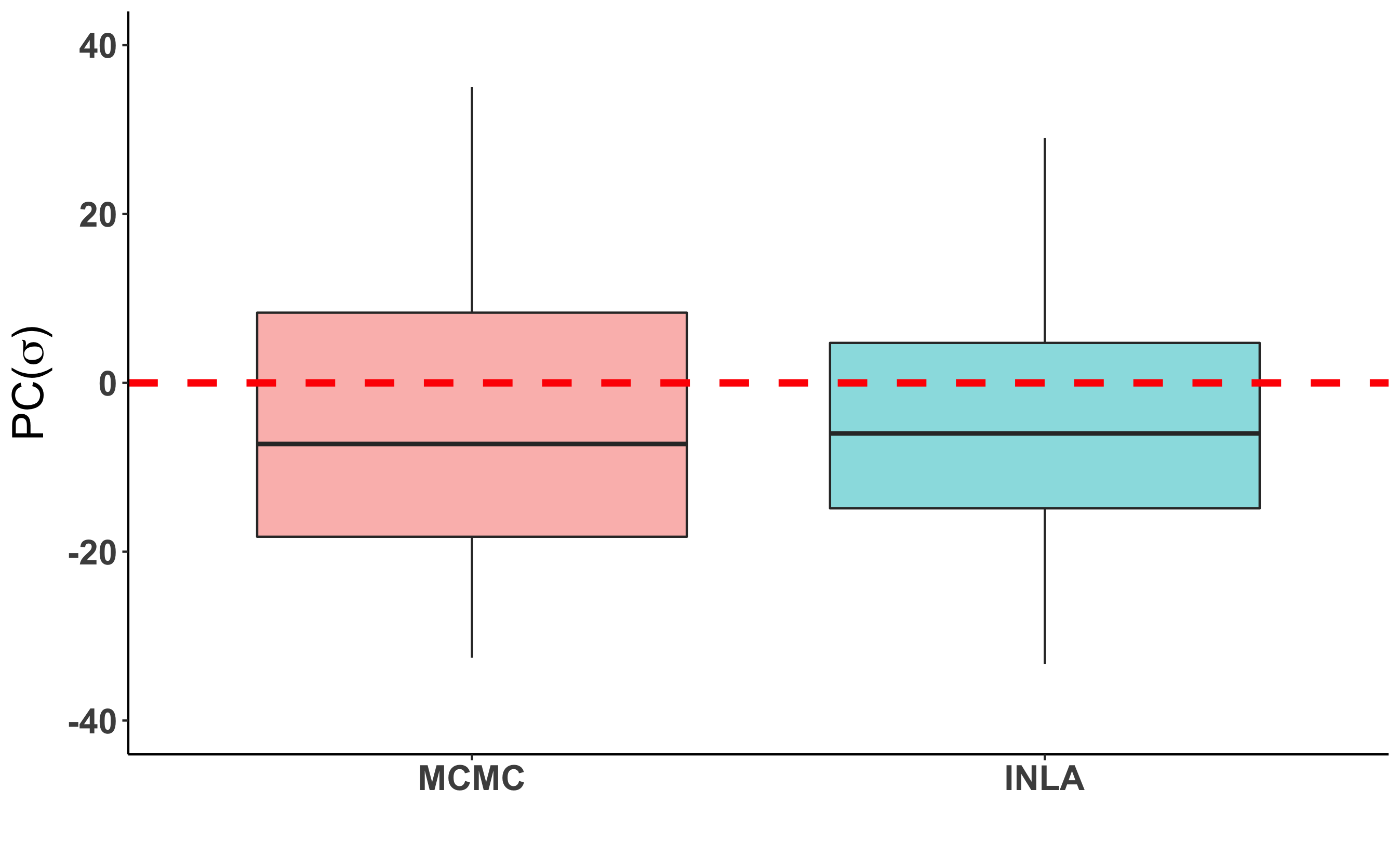}}
	%
	
	\caption[Description]{Boxplots of comparing posterior estimates of $\sigma$ for spatial data. (a) Percent error (PE). (b) Percent changes (PCs) from MCMC and INLA to the generating value 1.}
	\label{fig:bym_sigma}
\end{figure}

We first restrict our attention to $\beta_x$ and $\sigma$. As in \cref{sim:poisson}, we calculate $\textrm{PE}(\beta_x)$ and $\textrm{PE}(\sigma)$ for each dataset to assess how well INLA's output approximated the MCMC output. In \cref{fig:bym_beta}(a), we note that the majority of the time, the output from INLA still well approximated the output from MCMC for $\beta_x$. However, for 13 of 96 (14\%) of the datasets, $|\textrm{PE}(\beta_x)| > 30$ with one dataset having $\textrm{PE}(\beta_x)$ well over 100.  In \cref{fig:bym_sigma}(a), we see that INLA did not well approximate the output from MCMC for $\sigma$ in many of the datasets.  $|\textrm{PE}(\sigma)| > 30$ for 38 of the 96 (40\%) datasets INLA successfully fit.
We now compare the output from INLA and MCMC to the generating value. In \cref{fig:bym_beta}(b), the boxplots for $\textrm{PC}_{\textrm{MCMC}}(\beta_x)$ and $\textrm{PC}_{\textrm{INLA}}(\beta_x)$ showed similar distributions. However, in \cref{fig:bym_sigma}(b), the boxplots for $\textrm{PC}_{\textrm{MCMC}}(\sigma)$ and $\textrm{PC}_{\textrm{INLA}}(\sigma)$ showed more dissimilarities. It appears that, at least with respect to $\textrm{PC}_{\textrm{INLA}}(\sigma)$, INLA was actually a bit better at estimating the generating value of $\sigma$.

We now turn our attention to $\tau$. Both INLA and MCMC did a poor job of approximating this parameter. \citet{carroll2015comparing} had previously noted that the default settings in INLA resulted in poor inference for $\tau$ in a similar simulation study, as well.  Our results shed some light onto the potential factors driving this poor performance. The boxplots we have previously displayed are not meaningful given the extreme outliers. Instead, \cref{tab:bym_tau} summarizes the first three quantiles and the maximum values of estimates for $\tau$ (represented by $\textrm{E}(\tau | \bm{y})$ for each method) from INLA and MCMC. INLA would occasionally estimate it well but had very extreme outliers for some datasets; whereas MCMC appeared to consistently overestimate $\tau$. The MCMC performance suggests that the default priors in INLA are driving some of the poor performance. However, we note that for 41 of the 96 (43\%) datasets, $|\textrm{PE}(\tau)| > 30$, and for most of these datasets, INLA provided better estimates of $\tau$ than MCMC did. Therefore, it seems that something beyond the choice of priors was influencing INLA's performance.

Now, we consider what these results mean for \cref{quesone}. Because the estimates for $\tau$ included such large outliers, we restrict our attention to $\beta_x$ and $\sigma$ for a more meaningful discussion. Recall, there are currently two categories of accuracy measures in papers assessing INLA: comparison of INLA results to MCMC results ($\textrm{PE}(\cdot)$ for us) and comparisons of MCMC/INLA to a generating value ($\textrm{PC}_{INLA}(\cdot)$ and  $\textrm{PC}_{MCMC}(\cdot)$ for us).  For each parameter, we begin by restricting our attention to datasets for which INLA did not well approximate the results from MCMC.

\begin{figure}[h] 
	\centering
	\subcaptionbox{}{\includegraphics[width=0.45\textwidth]{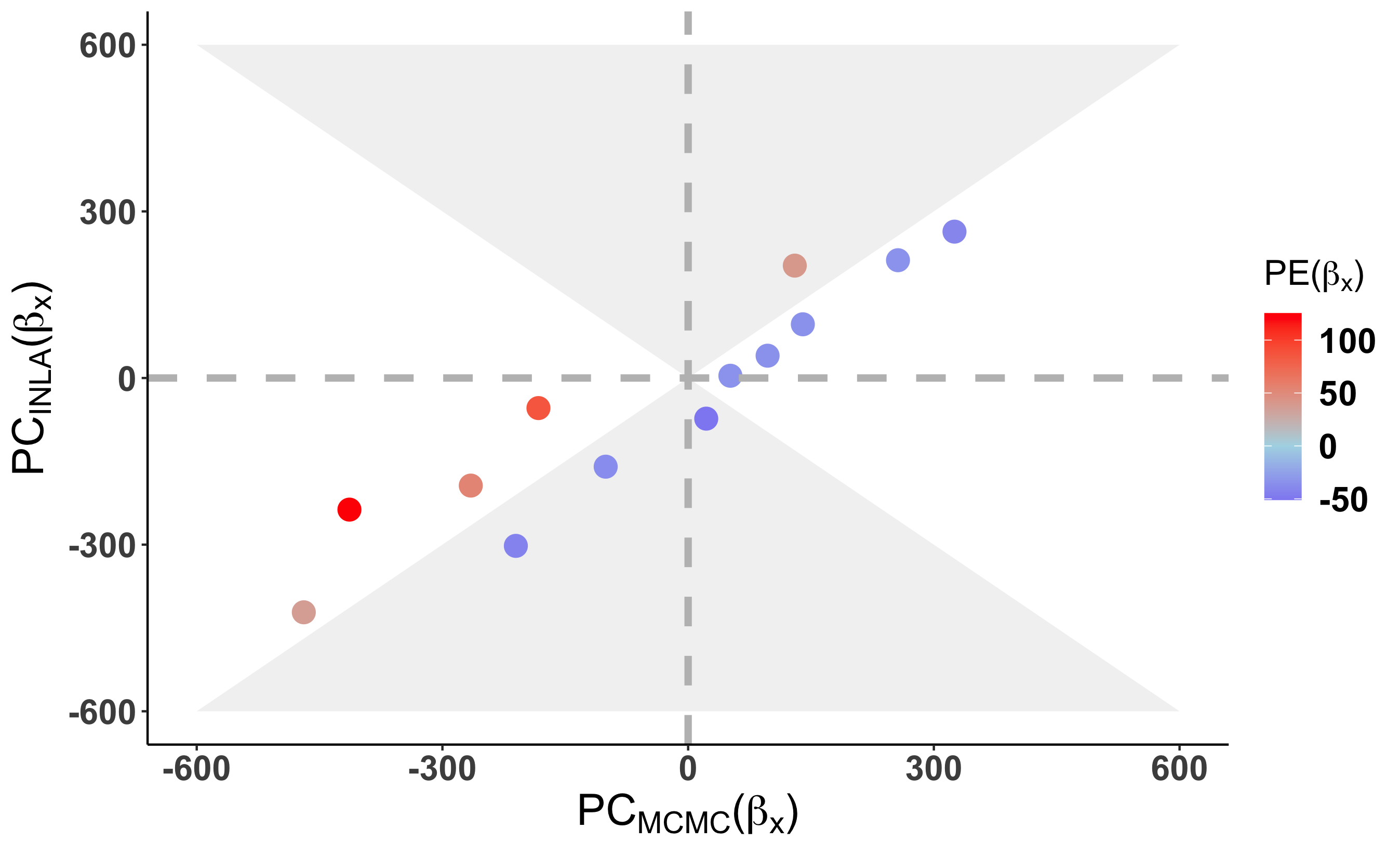}}
	\hfill 
	\subcaptionbox{}{\includegraphics[width=0.45\textwidth]{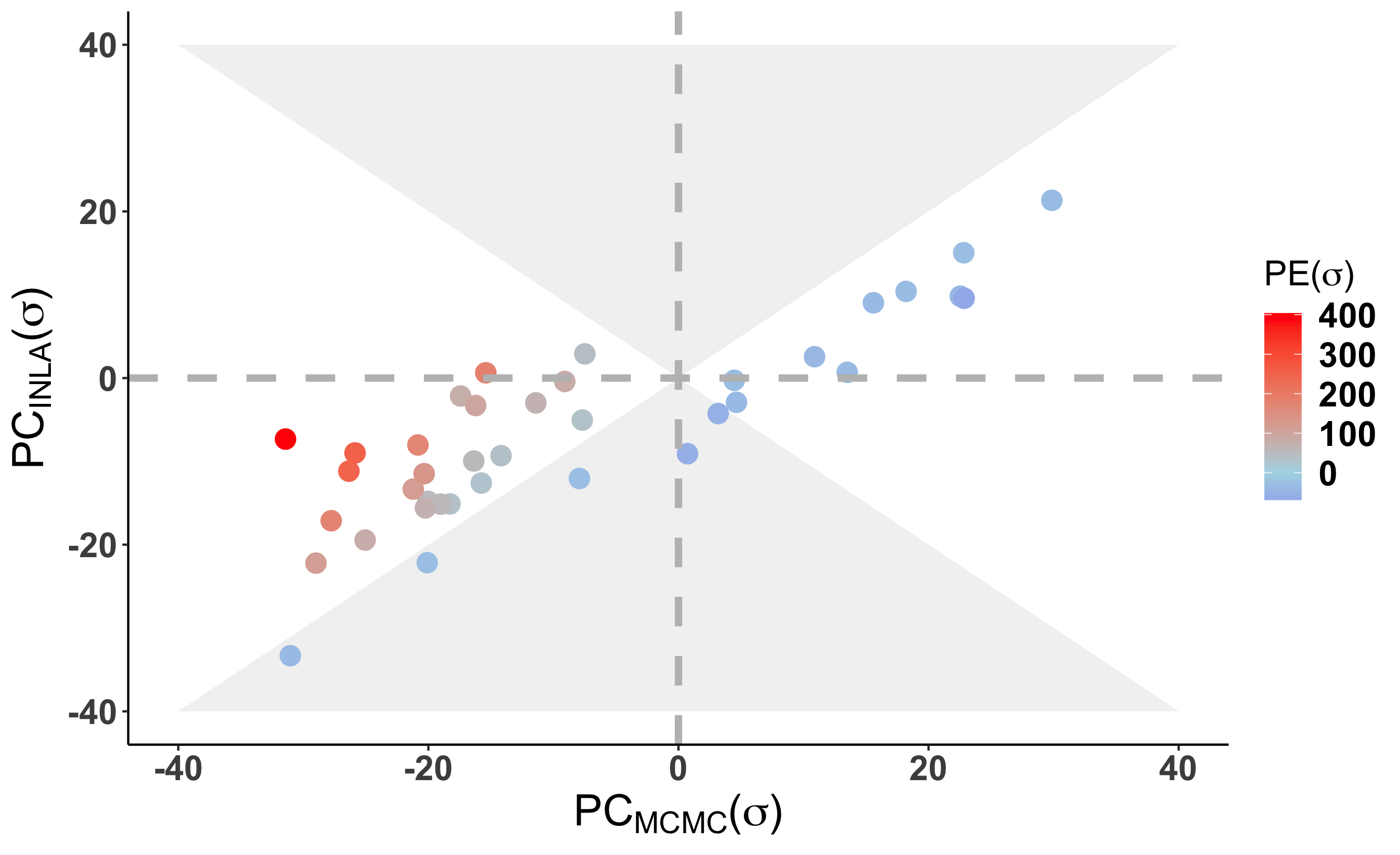}}%
	\hfill 
	\caption[Description]{These plots look at output datasets that produced significantly different results (PE $>$ 30\%) for INLA and MCMC for (a) $\beta_x$ and (b) $\sigma$. Note that points in the shaded gray area indicate datasets in which the MCMC estimate was closer to the generating value than the INLA estimate. }
	\label{fig:bym__compare}
\end{figure}

For datasets for which $|\textrm{PE}(\beta_x)| > 30$, \cref{fig:bym__compare}(a) plotted the  $\textrm{PC}_{\textrm{MCMC}}(\beta_x)$ against the  $\textrm{PC}_{\textrm{INLA}}(\beta_x)$. Similarly, for datasets for which $|\textrm{PE}(\sigma)| > 30$, \cref{fig:bym__compare}(b) plotted the  $\textrm{PC}_{\textrm{MCMC}}(\sigma)$ against the  $\textrm{PC}_{\textrm{INLA}}(\sigma)$. If the two categories of accuracy yielded consistent results, then we would expect to see all the points on these plots well away from the origin. However, for both $\sigma$ and $\beta_x$, there were a couple of points near the origin. Interestingly, for the majority of the datasets for which INLA and MCMC had very different estimates, INLA tended to produce results closer to the generating value.  Thus, with respect to \cref{quesone}, we found that comparisons of MCMC output to INLA output and comparisons of INLA/MCMC output to the generating value \textit{can} lead to very different conclusions at times. 

We were not able to identify any characteristics of the data that could help explain the more extreme values for $PE$ or the $PC$'s calculated, or when INLA/MCMC would yield estimates closer to the generating value. Our exploration was hindered, in part, by the fact that INLA has no diagnostics about how well (or poorly) the model fit (in a computational sense). This limitation seems to have been acknowledged in \citet{fong2010bayesian}, but has not been subsequently addressed to our knowledge. As an aside, in our explorations, we discovered that INLA would give estimates about the marginal posterior distributions for $\beta_x$, $\sigma$, and $\tau$ even if the marginal posterior distributions were not proper. For example, we fit a BYM model with an intercept and did not enforce a sum-to-zero constraint. The MCMC analysis resulted in trace plots that diverged, indicating nonconvergence. The INLA output simply gave us estimates in line with those obtained in the rest of our simulation study results.

We now turn our attention to \cref{questhree}. As in \cref{sim:poisson}, we began by attempting to reproduce the results from fitting the BYM model to one dataset. Again, we found we could only reproduce our results on the same machine and with the same version of INLA if we changed the number of threads to one. For the BYM model, unlike the Poisson regression model, both the version of INLA and the operating system could lead to drastically different results. For example, the same code for a BYM model including an intercept and a sum-to-zero constraint in INLA resulted in estimates of the intercept coefficient that differed by over $10^3$. When we repeated this exercise with NIMBLE, as before, we were able to obtain the same results for an MCMC analysis on two different operating systems and with two different versions of NIMBLE. 

This section only displays the results from one of the four combinations of INLA versions and operating systems (see \cref{data_code}).  We did rerun the code on the other three combinations, and we found that the results could vary significantly. For example, INLA failed to fit the BYM model to a different subset of the datasets for each of the three combinations. The number of datasets INLA failed to fit did not vary widely (INLA failed to fit a maximum of 6 datasets). For the most part, estimates of $\beta_x$ were somewhat stable. However, the estimates related to $\sigma$ and $\tau$ could be radically different. For example, on different operating systems, analyses run on the same version of INLA could result in estimates of $\textrm{E}(\tau | \bm{y})$ that differed by over $10^5$. The identifiability of the precision parameters $\tau$ and $\sigma$ is known to be problematic in the BYM model \citep{wakefield2007disease,roos2015sensitivity}. However, the erratic estimation of the marginal posterior distributions for $\tau$ and $\sigma$ is still concerning because these distributions are often used in practice to interpret relative risks of COVID-19 for areal units \citep{dimaggio2020black}.

\begin{table} 
\small\sf\centering
		\caption{Summaries of the means of the marginal posterior distributions for $\tau$. For INLA, only 96 datasets could be fit. For MCMC, the first number is the statistic from all 100 datasets, and the number in paranthases is the statistic from the 96 datasets with valid results from INLA.} 	\label{tab:bym_tau}
		\resizebox{\columnwidth}{!}{%
		\begin{tabular}{l cccc} 
			\toprule
			\multicolumn{1}{c}{Method} & \multicolumn{1}{c}{$Q_1$} & \multicolumn{1}{c}{$Q_2$} & \multicolumn{1}{c}{$Q_3$} & \multicolumn{1}{c}{Max}\\
			\midrule
			INLA 20.3.17 & 1.2 & 1.9 & 5.9 & 731,764.0 \\ 
			MCMC & 1.1 (1.14) & 2.6 (3.12) & 1667.8 (1703.8) & 2101.5 (2101.5) \\
			\bottomrule
		\end{tabular} }
\end{table}

\subsection{Model Selection} \label{sim:model selection}

In this section, we focus on \cref{questwo}. In our literature search, we found that many researchers analyzing COVID-19 data relied on INLA for model selection. Thus, in this section, we explore the use of INLA for model selection. There are several popular model selection criteria defined in R-INLA, including the Deviance Information Criterion (DIC) and Watanabe-Akaike information criterion (WAIC) \citep{gelman2014understanding,spiegelhalter2002bayesian}. The DIC calculated in R-INLA does not appear to be directly comparable to the versions calculated in most popular software \citep{carroll2015comparing}. 
However, a review of the documentation indicates that both NIMBLE and R-INLA use the same definition of the WAIC. Therefore, we use this criterion in the following simulations.

We first used the 200 datasets generated in \cref{sim:poisson,sim:bym}. During the simulation studies in \cref{sim:poisson,sim:bym}, we calculated the WAICs for each model we fit. We then fit a BYM model with the INLA and MCMC methods to the $100$ datasets generated from the Poisson regression model in \cref{sim:poisson}. Similarly, we used the $100$ datasets generated in \cref{sim:bym} and we fit a Poisson regression model. The BYM models and Poisson regression models were again fit using INLA's default priors and with the same methods described in \cref{sim:poisson,sim:bym}.

For WAIC, lower values are preferred. Therefore, for a given dataset, we considered INLA/MCMC to have ``selected'' the model with the lower WAIC. If the model selected matched the model the data was generated from, we considered this selection to be ``correct.'' In the following discussion, we refer to the data generated from the Poisson regression model as ``non-spatial'' data. Similarly, we refer to the data generated from the BYM model as ``spatial'' data.

\paragraph{Non-Spatial Data}

 When we used INLA to fit a Poisson model to each of the $100$  non-spatial datasets, all models were successfully fit. However, when we used INLA to fit a BYM model to each of the datasets, one of the models failed to fit.  For the MCMC method, we were able to fit all 100 datasets. In this section, we used 99 datasets which were successfully fit to both the BYM and Poisson models when comparing the results obtained from INLA and MCMC. 
 
\begin{figure}[H] 
	\centering
	\includegraphics[width=0.60\textwidth]{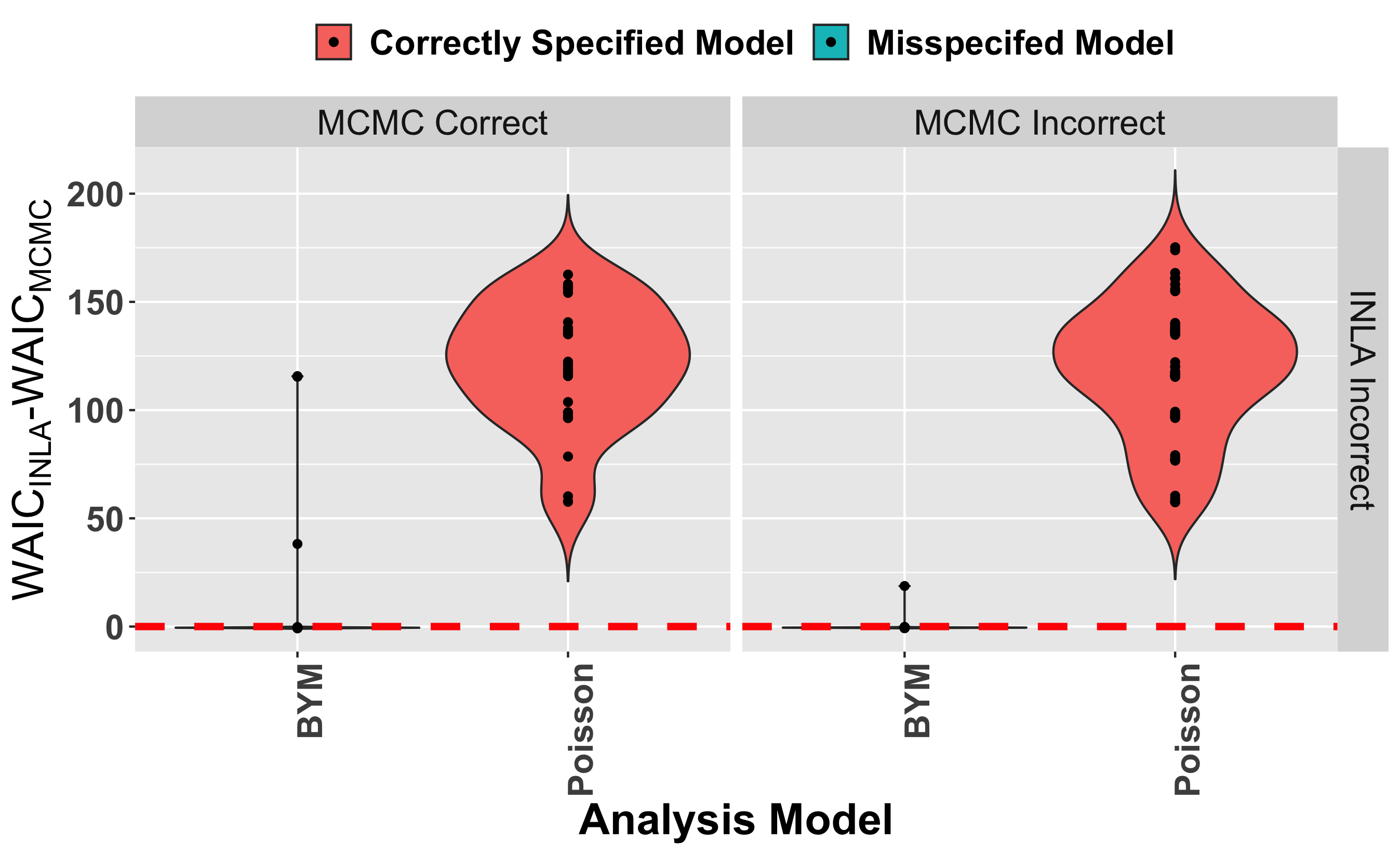}
	\caption[Description]{These plots look at the $\textrm{WAIC}_{INLA} - \textrm{WAIC}_{MCMC}$. The $x$-axis indicates the model that was fit. The color red indicates the scenario that the data were generated from the analysis model. The violin plots in the left panel were generated from 54 datasets. The violin plots in the right panel were generated from 45 datasets.}
	\label{fig:ns_model}
\end{figure}

 For these 99 datasets, INLA never selected the correct model. MCMC, on the other hand, selected the correct model 54 times. In order to understand what was driving the differences in model selection between the analyses by MCMC and INLA, we considered the following quantity: $\textrm{WAIC}_{INLA} - \textrm{WAIC}_{MCMC}$. In \cref{fig:ns_model}, we considered the distributions of this quantity based on whether an incorrect or correct selection was made by each method. Recall that in \cref{sim:poisson}, INLA seemed to approximate the output from MCMC quite well for the Poisson regression model. However, in \cref{sim:bym}, we found that INLA was not as reliable in approximating the output from MCMC for the BYM model. Surprisingly, here, we found a reverse of this pattern for the calculation of the WAIC. INLA did not do a particularly good job at approximating the WAIC that was obtained by MCMC for the Poisson regression model. Instead, the WAIC calculated by INLA was systematically inflated compared to the WAIC calculated from the MCMC method. However, for the BYM model, which was misspecified for these datasets, the WAIC from INLA tended to match the WAIC from MCMC.

\paragraph{Spatial Data}

INLA successfully fit the Poisson model for all 100 of the spatial datasets, and it successfully fit the BYM model for 96 datasets. Thus, all comparisons between INLA and MCMC involved only 96 datasets.

For the 96 datasets it successfully fit, INLA selected the correct model for 40 (42\%) of them. For the 100 datasets successfully fit with MCMC, MCMC selected the correct model for 40 (40\%) of them. As before, we compared the WAIC obtained from the Poisson model and the BYM for the 96 datasets. INLA selected the same model as MCMC for 51 (53\%) of the datasets. However, of these 51 datasets, the selection was correct only 16 times. In other words, the cases for which the INLA and MCMC model selections agreed were mostly cases in which the incorrect model was selected.

\begin{figure}[H] 
	\centering
	\includegraphics[width=0.60\textwidth]{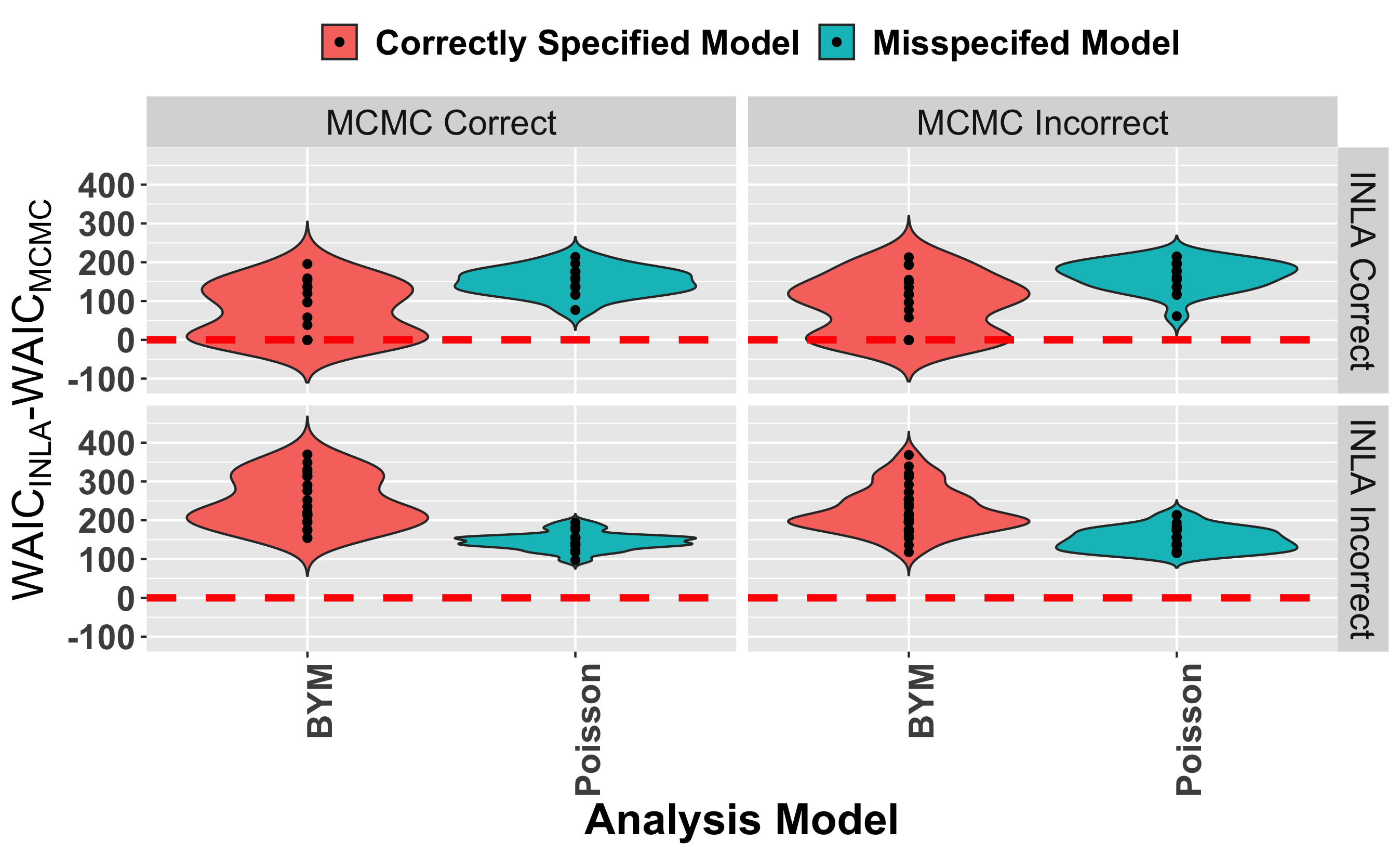}
	\caption[Description]{ These plots look at the $\textrm{WAIC}_{INLA} - \textrm{WAIC}_{MCMC}$ for the 96 datasets fit successfully by both INLA and MCMC. The $x$-axis indicates the model that was fit. Starting from the upper left hand corner and moving clockwise, the violin plots in each panel were generated from $n=16$, $n=24$,$n=35$, and $n=21$ datasets, respectively.   }
	\label{fig:sp__model}
\end{figure}

We again considered the difference in WAIC calculated, i.e., $\textrm{WAIC}_{INLA} - \textrm{WAIC}_{MCMC}$ based on the model selection made by each method. In \cref{fig:sp__model}, we note that, compared to the WAIC computed by MCMC, INLA systematically inflated the WAIC for Poisson models for all 4 scenarios. Unlike in \cref{fig:ns_model}, however, the WAIC from INLA did not match the WAIC from MCMC for the BYM model. Instead, it too tended to be inflated. The datasets for which this inflation tended to be greater were the datasets that INLA tended to select the incorrect model.

Taken together with our results from \cref{sim:poisson,sim:bym}, the results in this section suggested that traditional methods of assessing the accuracy of INLA's approximations were inadequate in this setting. 
We note that the WAIC's from INLA tend to be less reproducible than the summaries of the marginal posterior distributions considered. Specifically, for both the Poisson regression model and the BYM model, the WAIC's changed when the operating system changed (holding the version of INLA fixed) or the version of INLA changed (holding the operating system fixed). 

\section{Case Study of COVID-19 Data} \label{sec:casestudies}

In this section, we use INLA to reproduce one of the analyses of COVID-19 data completed in \citet{millett2020assessing}, in which the authors used data on COVID-19 cases in $3,108$ U.S. counties to assess racial disparities. In one of their analyses, they used INLA to fit a zero-inflated negative binomial model to their data (see Table 2 of \citet{millett2020assessing} for their results). We refrain from describing their data in detail here, but refer the interested reader to \citet{millett2020assessing}. In reproducing this analysis, we aimed to answer two questions: 1)  Would the authors have reached the same conclusions if they had used MCMC to fit the model instead? and 2) Are their results reproducible?

Logistically, we ran into problems attempting to assess these two questions simultaneously. The model used in \citet{millett2020assessing} included a state-specific random effect. However, we found that including such a term for MCMC lead to trace plots with trends (indicating convergence had not been reached). In order to compare INLA to MCMC, we omitted the state-specific random effect in this section. We also ran the exact code from \citet{millett2020assessing} to more thoroughly explore reproducibility concerns (for more details, see \appendixnegbin).

According to the R-INLA documentation, the zero-inflated negative binomial model can be written as follows: Assume that $y_i$ are the number of COVID cases in county $i$, for $i= 1, \ldots, 3108$, then for $j= 1,2,3, \ldots$:
\[  \textrm{P}\left( y_i = j | p_z, \mu_i, n \right)  = p_z \times 1_{[y_i=0]}   + \left( 1 - p_z \right) \textrm{NB} \left(y_i \right), \]
where $p_z$ is the probability that $y_i$ is $0$, for $i= 1, \ldots, 3108$ and  $\textrm{NB}$ represents the negative binomial likelihood parameterized with mean $\mu_i$ and dispersion parameter $n$. The parameter $\mu_i$ is related to the covariates and an offset through the log link-function:
\[ \log (\mu_i) = \beta_0 + \beta_1 x_{1i} + \ldots + \beta_k x_{ki}  + \log( \textrm{Population}_i). \]
The value $\textrm{Population}_i$ is the population for county $i$. We employed the default prior specifications from INLA with one exception: The regression coefficients were given normal priors with mean $0$ and standard deviation $1000$. Instead of specifying priors on $p_z$ and $n$, INLA specifies priors on $\theta_1 = \textrm{logit}(p_z)$ and $\theta_2 = \log(n)$. The former is given a normal prior with mean -1 and precision .2; the latter is given what appears to be a member of the penalized complexity priors \citep{simpson2017penalising,fuglstad2019constructing}. Based on the current documentation, it does not appear that this particular prior had been evaluated in peer-reviewed literature at the time that \citet{millett2020assessing} was written. Because one of the purposes of this assessment is to see how closely INLA output approximates MCMC output, we decided to use a different prior available in INLA: a flat prior on $\theta_2 = \log(n)$. 

As in \cref{sec:simulation}, we used NIMBLE to ensure that the model fit with MCMC was parameterized exactly as it is in the R-INLA documentation. To ensure that INLA and MCMC obtained the same information, we used complete cases for each of the 15 covariates included in \cref{tab:negbinomcompare} (See \appendixnegbin ~ for more details). For MCMC, the algorithm was run 2,000,000 times and a burn-in of 100,000 was used. Convergence was assessed with trace plots for all parameters.

In \cref{tab:negbinomcompare}, we computed the modifiable rate ratios defined by \citet{millett2020assessing} based on the output from the INLA and the MCMC analysis. In many fields, including epidemiology, it is common to identify covariates associated with rate ratios' whose 95\% credible interval excludes 1 as important (or ``significant'') predictors of the response of interest. Rate ratios whose 95\% credible interval include only values above $1$ are considered positively associated with COVID-19 cases, and those whose 95\% credible intervals include only values below $1$ are considered negatively associated with COVID-19 cases. This is, in fact, what \citet{millett2020assessing} did, and so, it is what we do here.

In \cref{tab:negbinomcompare}, we see that, except for the days since the first case of diagnosis, the modifiable rate ratios computed by MCMC and R-INLA did not seem very different. However, our inference based on the credible intervals would have been fairly different. When we fit the model with MCMC, we would have found 8 covariates associated with COVID-19 cases. We would have only derived the same conclusion as the results from INLA for 5 of them (i.e., percent Black, HIV infection rate, air toxins, social distancing, and days since the first diagnosis). Both the INLA and MCMC outputs indicated a significant relationship with percent uninsured; however, they disagreed with the direction of this relationship. 

We now turn our attention to the question of reproducibility. When we ran the exact code from \citet{millett2020assessing}, we found various factors that could impact the output. Some of them include which functions in INLA were used to calculate the modifiable rate ratios, the operating system, and the version of INLA used. In \appendixnegbin, we explain in more detail how the version of INLA and the method of calculating rate ratios in INLA might impact differences in inference. To summarize, for this model, the version of INLA and the INLA functions used can impact the conclusions reached by this model, although it only changed inference for one covariate. However, we did not find the operating system had any impact on the conclusions reached.

In this paper, we have defined reproducibility very narrowly. However, the changes we have made to the model ran in \citet{millett2020assessing} and the model ran here are small enough that they may not have been explicitly stated in some applied papers. So, we find it meaningful to take a moment to compare our output here to the output obtained in \citet{millett2020assessing}. The authors found that 5 of the 15 covariates listed in \cref{tab:negbinomcompare} were associated with the rates of COVID-19 cases in a county. In particular, they concluded that higher rates of COVID-19 cases were associated with a higher proportion of black residents, a higher proportion of uninsured residents, a higher proportion of residents living in a crowded environment, and more days since the first case in a county. Additionally, they found that higher social distancing scores (i.e., poor social distancing practice) were associated with lower rates of COVID-19 cases.

Referring to \cref{tab:negbinomcompare}, we see that our version of the negative binomial model would have found 8 of the 15 covariates were associated with the number of COVID-19 cases in a county. Only four of these covariates were shared with those found in Table 2 of \citet{millett2020assessing}. There are a few factors that could be contributing to this difference. The first, of course, is the differences in the model specifications: the state-specific random effect and the prior on $\log(n)$. As already noted, we found that including a state-specific random effect led to convergence issues. However, it is difficult to investigate the impact of the change of the prior. \citet{millett2020assessing} did not report the version of 
INLA they used. According to the R-INLA website, the INLA documentation is kept up to date with the current testing version of the package (updated weekly). Therefore, we have no way of knowing the default prior in use at the time of the initial analysis. From a reproducibility stand point, this makes it very difficult to ever formally assess whether a change in priors is impacting inference.

To summarize, in this section we found that the conclusions reached based on INLA output did not match the conclusions reached based on output from MCMC for the zero-inflated negative binomial data examined here. For one of the covariates (percent uninsured), MCMC and INLA would have led researchers to conclude different directions of associations between it and the number of COVID-19 cases in a county. From a reproducibility standpoint, we found that the version of INLA could sometimes lead to different conclusions about the importance of a covariate.

\begin{table}
\small\sf\centering
\caption{Adjusted rate ratios (third versus first quartile) as defined in \citet{millett2020assessing}. Rate ratios greater than one mean that higher levels of a given characteristic are associated with higher rates of COVID-19 cases.  Rate ratios lower than one mean that higher levels of a given characteristic are associated with lower rates of COVID-19 cases. \newline * indicates a 95\% credible interval excluding one.} 
\nopagebreak
\resizebox{\columnwidth}{!}{%
\begin{tabular}{l l l} 
\toprule
\multicolumn{1}{c}{Covariate} & \multicolumn{1}{c}{Method} & \multicolumn{1}{c}{Rate Ratio (95\% Cred. Int.)}\\
\midrule
\multirow{2}{*}{Percent Black} 
 & INLA 20.3.17 & 1.194 (1.171, 1.262)*\\ 
 & MCMC & 1.108 (1.047, 1.171)*\\
\midrule
\multirow{2}{*}{Percent White} 
 & INLA 20.3.17 & 0.983 (0.950, 1.080) \\ 
 & MCMC & 0.887 (0.810, 0.972)* \\
\midrule
\multirow{2}{*}{Percent 65 +}
 & INLA 20.3.17 & 0.951 (0.930, 1.012) \\ 
 & MCMC & 0.945 (0.888, 1.007) \\
\midrule
\multirow{2}{*}{Percent unemployed} 
 & INLA 20.3.17 & 0.920 (0.903, 0.972)*  \\ 
 & MCMC & 0.977 (0.922, 1.034) \\
\midrule
\multirow{2}{*}{Percent uninsured} 
 & INLA 20.3.17 & 1.029 (1.006, 1.095)* \\
 & MCMC & 0.931 (0.872, 0.992)* \\
\midrule
\multirow{2}{*}{Percent diabetes diagnoses} 
 & INLA 20.3.17 & 0.974 (0.955, 1.029) \\ 
 & MCMC & 0.936 (0.883, 0.993)* \\
\midrule
\multirow{2}{*}{Heart disease death rate} 
 & INLA 20.3.17 & 1.085 (1.064, 1.148)* \\ 
 & MCMC & 1.042 (0.981, 1.104) \\
\midrule
\multirow{2}{*}{HIV infection rate} 
 & INLA 20.3.17 & 1.032 (1.018, 1.072)* \\ 
 & MCMC &  1.058 (1.016, 1.102)* \\
\midrule
\multirow{2}{*}{Cerebrovascular and hypertension death rate}
 & INLA 20.3.17 & 1.025 (1.012, 1.060)* \\ 
 & MCMC & 1.034 (0.989, 1.080) \\
\midrule
\multirow{2}{*}{Urbanicity score} 
 & INLA 20.3.17 & 1.009 (0.976, 1.104) \\ 
 & MCMC & 0.936 (0.849, 1.030) \\
\midrule
\multirow{2}{*}{Air toxins (PM$_{2.5}$)} 
 & INLA 20.3.17 &  1.151 (1.120, 1.241)*  \\ 
 & MCMC & 1.186 (1.095, 1.279)* \\
\midrule
\multirow{2}{*}{Household occupancy >1 person per room} 
 & INLA 20.3.17 &  1.003 (0.988, 1.049) \\ 
 & MCMC & 1.023 (0.977, 1.072) \\
\midrule
\multirow{2}{*}{Social distancing score} 
 & INLA 20.3.17 & 0.786 (0.773, 0.823)* \\ 
 & MCMC & 0.751 (0.715, 0.788)*  \\
\midrule
\multirow{2}{*}{Days since the first case of diagnosis} 
 & INLA 20.3.17 & 3.639 (3.544, 3.917)* \\ 
 & MCMC & 1.978 (1.831, 2.129)* \\
\bottomrule
\end{tabular} }
\label{tab:negbinomcompare}
\end{table}

\section{Discussion} \label{sec:discussion}
In this paper, we were motivated by an important, but yet unexplored, question: Are the analyses of COVID-19 data conducted with R-INLA providing us with accurate and reproducible results? To answer this question, we first explored how the accuracy of INLA's approximations have been previously evaluated. For most settings, there are no theoretical guarantees about these approximations. Instead, INLA's approximations have largely been evaluated with illustrative examples and case studies. The measures of accuracy used in these settings have not been consistent. Furthermore, many applied researchers use INLA in ways which are not supported by existing methods of assessing the accuracy of INLA's approximations. Finally, despite the fact that INLA is primarily validated by using the R-INLA package to analyze case studies, there do not appear to have been previous attempts to assess the reproducibility of analyses conducted with INLA. We explored each of these limitations with simulation studies and an attempt to reproduce the results of peer-reviewed article. We ultimately found that these limitations could change the conclusions researchers drew about COVID-19.

This paper has focused on the use of INLA to fit disease mapping models to COVID-19 data. In many ways, this is just a special case of a broader problem. More and more complex statistical methods are being proposed in statistical literature. Many of these methods lack the theoretical guarantees of their historical counterparts and rely on computationally advanced implementations. How then, can medical researchers ensure that the use of these new techniques result in accurate and reproducible results? To our knowledge, no one has attempted to tackle this problem before. Thus, we take a moment to use the insights we have gained in our exploration of INLA to propose a minimal set of guidelines when using statistical methodology that primarily relies on simulation studies or case studies to validate its use. We express these points generally, but we also provide context with our findings about INLA.

\begin{enumerate}
    \item \textbf{The statistical methodology should result in reproducible analyses.} 
\end{enumerate}
Recall, we found that the default settings in the R-INLA package prevent even the narrow reproducibility defined in \cref{reproduce}. However, every paper we reviewed (for which code was available) left these default settings in place. This was true both for research seeking to validate the use of R-INLA and for applied research. In practice, this means that another researcher might not be able to re-run the code and obtain the same results.

If a statistical methodology has been validated with case studies or simulation studies, we propose that those studies should be reproducible (at least in the sense defined in \cref{reproduce}). Additionally, we suggest that the analyses ran by medical researchers should be reproducible as well. 

\begin{enumerate}
\setcounter{enumi}{1}
    \item \textbf{The statistical methodology should have been \textit{meaningfully} assessed for the type of data, models, and goals of analysis under consideration.}
\end{enumerate}
We found that many applied researchers use INLA in ways that are not  supported by existing methods of assessing the accuracy of INLA's approximations. For example, many researchers used INLA's WAIC to perform model selection. However, when we attempted to use INLA to select between two models with WAIC, we found INLA did not perform well. The performance depended not only on the model fit, but also on the data used to fit it. For example, the WAIC from INLA closely matched the WAIC from MCMC when the BYM model was fit to non-spatial data, but this did not remain true when the BYM model was fit to spatial data. 

If a statistical methodology has primarily been validated with simulation studies or case studies, it is important that medical researchers do not venture into a setting where the methodology has not been tested. It is possible, just as we saw with INLA, that the methodology may not perform well in this new setting. We propose that a \textit{meaningful} assessment should evaluate the methodology's ability to compute quantities that the applied researcher wishes to use for more than one dataset.

\begin{enumerate}
\setcounter{enumi}{2}
    \item \textbf{The statistical methodology should  be capable of producing statistical diagnostic measures.}
\end{enumerate}
Unlike the other two suggestions, this suggestion will depend heavily on the type of statistical methodology under consideration. So, we use our experience with INLA as an illustrative example. Recall, we found that INLA would produce results even when the posterior distribution under consideration was improper. There was no obvious indication of the fact that the model should not have fit. In this setting, a fully Bayesian approach would have indicated the problem via various diagnostic statistics designed to assess convergence. Ideally, INLA would be compatible with a diagnostic that indicated the impropriety as well.

\newpage

\noindent \textbf{Supplementary materials}
Supplementary materials include results obtained from re-running the code provided with \citet{millett2020assessing} on 4 combinations of operating system and R-INLA package version.

\noindent \textbf{Acknowledgments}
HL was supported by the Director, Office of Science, of the U.S. Department of Energy under Contract No. DE-AC02-05CH11231.

\bibliographystyle{unsrtnat}
\bibliography{references}

\newpage
\appendix
\section{Code and Data} \label{data_code}

\paragraph{Code and R Data Objects}
The original data, simulated data, code, and output used to generate the plots and images in this paper can be downloaded as a zip file at \url{https://korikhan.com/code-and-data/}. The INLA documentation relied on in this paper has also been saved in the above-mentioned folder. 

The analyses summarized in \cref{sec:simulation} were run on Windows 10 x64 (referred to as WINDOWS in \appendixnegbin) with R version 4.0.3, INLA version 21.02.23, and NIMBLE version  0.11.1. The analyses reported in \cref{sec:casestudies} were run on macOS Catalina 10.15.7 (referred to as MAC in \appendixnegbin) with R version 4.0.2, INLA version 20.03.17, and NIMBLE version  0.9.1. All plots and images were generated with the ggplot2 package version 3.3.3 \citep{ggplot2}. 

\paragraph{Data Sources}

The data used to generate the datasets in \cref{sim:poisson}--\cref{sim:model selection} came from two sources. The counts of COVID-19 tests were collected from \url{https://dhsgis.wi.gov} on January 20, 2021. The poverty rates were obtained from \url{https://data.census.gov} on January 26, 2021. The data analyzed in \cref{sec:casestudies} was originally downloaded from \url{https://github.com/benkeser/covid-and-race}.

\end{document}